\documentclass[acmsmall]{acmart}

\usepackage{array}
\usepackage{breqn}
\usepackage{enumitem}
\usepackage{tikz,pgfplots}
\usepackage{booktabs,makecell,multirow}
\usetikzlibrary{positioning}
\usetikzlibrary{arrows}
\usetikzlibrary{patterns}
\usetikzlibrary{shapes.geometric}
\usetikzlibrary{decorations.pathreplacing}
\usetikzlibrary{calligraphy}
\usetikzlibrary{hobby}
\usetikzlibrary{calc}

\usepackage{pifont}
\newcommand{\cmark}{\ding{51}}
\newcommand{\xmark}{\ding{55}}

\newtheorem{theorem}{Theorem}[section]

\graphicspath{{./figures/}}

\AtBeginDocument{%
  }

\begin{document}

\title{Memento Filter: A Fast, Dynamic, and Robust Range Filter}

\author{Navid Eslami}
\email{navideslami@cs.toronto.edu}
\affiliation{%
  \institution{University of Toronto}
  \city{Toronto}
  \country{Canada}
}

\author{Niv Dayan}
\email{nivdayan@cs.toronto.edu}
\affiliation{%
  \institution{University of Toronto}
  \city{Toronto}
  \country{Canada}
}


\begin{abstract}
    Range filters are probabilistic data structures that answer approximate
    range emptiness queries. They aid in avoiding processing empty range
    queries and have use cases in many application domains such as key-value
    stores and social web analytics. However, current range filters do not
    support dynamically changing and growing datasets. Moreover, several of
    these designs also exhibit impractically high false positive rates under
    correlated workloads, which are common in practice. These impediments
    restrict the applicability of range filters across a wide range of use
    cases.

    We introduce Memento filter, the first range filter to simultaneously offer
    dynamicity, fast operations, and a robust false positive rate for any
    workload. Memento filter partitions the key universe and clusters its keys
    according to this partitioning. For each cluster, it stores a fingerprint
    and a list of key suffixes contiguously. The encoding of these lists makes
    them amenable to existing dynamic filter structures. Due to the one-to-one
    mapping from keys to suffixes, Memento filter supports inserts and deletes
    and can even expand to accommodate a growing dataset.

    We implement Memento filter on top of a Rank-and-Select Quotient filter and
    InfiniFilter and demonstrate that it achieves a competitive false positive
    rate and performance with the state of the art while also providing
    dynamicity. Due to its dynamicity, Memento filter is the first range filter
    applicable to B-Trees. We showcase this by integrating Memento filter into
    WiredTiger, a B-Tree-based key-value store, significantly boosting its
    performance for mixed workloads.
\end{abstract}

\setcopyright{acmlicensed}
\acmJournal{PACMMOD} 
\acmYear{2024} 
\acmVolume{2} 
\acmNumber{6 (SIGMOD)} 
\acmArticle{244} 
\acmMonth{12}
\acmDOI{10.1145/3698820}

\ccsdesc[500]{Theory of computation~Bloom filters and hashing}
\ccsdesc[500]{Information systems~Unidimensional range search}

\keywords{Range Filter; Dynamic Data Structure; Data Growth; Scalability}

\received{April 2024}
\received[revised]{July 2024}
\received[accepted]{August 2024}

\maketitle

\section{Introduction}
\textbf{What is a Filter?}
A filter is a compact probabilistic data structure that answers approximate
membership queries on a set. Since a filter is space efficient, it is often
stored in a higher level of the memory hierarchy, making it fast to query. A
filter cannot return a false negative but may return a false positive with some
probability known as the false positive rate (FPR), determined by its memory
footprint. Filters are ubiquitously used in many application domains to avoid
disk reads~\cite{RocksDB} or network hops~\cite{FiltersNetworking} when
querying for non-existing keys.

\textbf{Range Filters and Applications.}
Traditional filters, such as Bloom filters~\cite{Bloom}, only answer membership
queries for a single query key. A range filter, on the other hand, is a filter
that answers range emptiness queries over a set $S$~\cite{BeyondBloomTutorial}.
Given a range $q = [q_l, q_r]$, a range filter returns a true positive if there
is some key in $S$ that is also in the range $q$. It returns either a true
negative or a false positive otherwise. Range filters are used to avoid
processing empty range queries in applications such as social web
analytics~\cite{YCSB}, replication in distributed key-value stores~\cite{Rose},
statistics aggregation of time series~\cite{VarLenTimeSeries}, and SQL table
accesses~\cite{SQLTableAccesses}. Previous research has demonstrated the
significant performance boost range filters provide in these
systems~\cite{AdaptiveRangeFilter, SuRF, Rosetta, bloomRF, Proteus, SNARF,
Oasis}.

\textbf{The Need for Dynamic Range Filters.}
A static range filter is sufficient for applications with immutable (i.e.,
non-changing) data. For example, LSM-Trees consist of a set of immutable files
to which range filters can be attached to speed up range queries. However, many
applications have a dynamic nature, as they must support insertions, deletions,
and growing datasets. These include (1)~LSM-Trees that use a single global
filter to map each key to the file storing it~\cite{SlimDB, SplinterDB, Chucky,
GRF}, (2)~B-Tree indexes~\cite{BTree, B+Tree}, and (3)~Hybrid
Transactional/Analytical Processing (HTAP) systems~\cite{HTAPArticle,
HTAPBook}. Such applications require dynamic range filters that maintain high
performance and low FPRs with an increasing dataset size.

\textbf{State of the Art.}
All existing range filters~\cite{SuRF, Rosetta, REncoder, REncoder_Journal,
bloomRF, Proteus, SNARF, Oasis, Grafite} resemble Bloom filters and/or create a
static model of the data distribution. The Bloom filter-inspired
methods~\cite{Rosetta, REncoder, REncoder_Journal, bloomRF, Proteus, SNARF,
Oasis, Grafite} hash a key to one or more bits in a bitmap and set those bits
to ones, potentially mapping multiple keys to the same bit. These filters fail
to support deletions, as changing a bit from one to zero may result in false
negative query results. Furthermore, such filters are unexpandable as they
provide no obvious way of remapping keys to a larger
bitmap~\cite{InfiniFilter}. Those range filters that model the data
distribution~\cite{SuRF, REncoder, REncoder_Journal, Proteus, SNARF, Oasis}
also cannot update their models due to their static layout and the lack of
global information about the dataset. Thus, none of the current range filters
are dynamic or expandable.

Moreover, many of the above range filters \cite{SuRF, REncoder,
REncoder_Journal, Proteus, SNARF, Oasis} provide no FPR guarantees under common
workloads~\cite{Rosetta, Grafite} where the query end-points are close to the
keys in the set. For example, queries for employee salaries tend to target
ranges near the data values rather than unrealistic ranges. The FPR measured in
practice may approach 1, rendering the filter useless~\cite{Grafite}.

\textbf{Research Challenge.}
We identify the following research question: \emph{can we design a range filter
that simultaneously (1)~guarantees a theoretically optimal FPR for any key set
and workload, (2)~provides fast operations, and (3)~supports insertions,
deletions, and resizability while maintaining~(1)~and~(2)?}

\textbf{Core Contributions.}
We introduce Memento filter, the first range filter providing dynamicity,
resizability, fast operations, and an optimal FPR for any workload. It achieves
this by taking a new approach to range filtering. Memento filter partitions the
key space into equi-width partitions. It then clusters the keys according to
this partitioning and maintains a suffix for each key in a Rank-and-Select
Quotient filter (RSQF)~\cite{GQF}. It further derives a fingerprint for each
cluster based on its partition number. All suffixes in a given partition are
stored compactly and contiguously alongside their fingerprint to support
cache-efficient parsing. An RSQF employs Robin Hood
Hashing~\cite{RobinHoodHashing} to resolve hash collisions, allowing for
storing the variable length data of the partitions by pushing colliding filter
content to the right. As Memento filter establishes an unambiguous one-to-one
mapping from each key to a suffix (i.e., unlike Bloom filter-inspired
approaches), it can expand efficiently by remapping these suffixes to a larger
filter. Memento filter handles a range query by finding the partitions that
intersect the query range. It then compares the key suffixes of those
partitions and the suffixes of the query end-points to check if any fall within
the specified range.

\textbf{Additional Contributions:}
\vspace{-14pt}
\begin{itemize}
    \item We provide the most comprehensive theoretical comparison of range
        filters to date and show that Memento filter not only matches the state
        of the art in terms of FPR and performance but also provides dynamicity
        and expandability. 
    \item We use variable-length fingerprints to create an expandable variant
        of Memento filter with an optimal and robust FPR, similarly to
        InfiniFilter and Aleph Filter, among other expandable
        filters~\cite{InfiniFilter, AlephFilter, TaffyFilters}.
    \item We empirically evaluate Memento filter against all major range
        filters in a static setting. We also conduct the first evaluation of
        range filters in a dynamic setting.
    \item We integrate Memento filter with a B-Tree-based key-value store and
        show that it significantly boosts throughput for dynamic mixed
        workloads. Memento filter is the first range filter to achieve such a
        feat.
\end{itemize}

\section{Problem Analysis}
This section defines the problem of range filtering and shows that no existing
solution simultaneously provides (1)~a robust FPR for any workload, (2)~fast
worst-case performance, and (3)~the ability to handle dynamic data.

\textbf{Definitions and the Theoretical Lower Bound.}
The problem of range filtering is a generalization of the classic filtering
problem. A query takes the form of an interval $q = [q_l, q_r]$ of length at
most~$R$. Given a set of keys $S$ coming from a universe of size $u$, the goal
is to ascertain the emptiness of the range, i.e., whether or not $S \cap q \neq
\emptyset$ with an FPR of at most~$\epsilon$. The top section of
Table~\ref{tab:term_and_symbol_definitions} outlines the terms used to describe
the range filtering problem henceforth.

\begin{table}
    \centering
    \smaller
    \begin{tabular}{cl}
        \toprule 
        \textbf{Symbol} & \textbf{Definition} \\
        \midrule
        $q = [q_l, q_r]$ & An inclusive query range. \\
        $u$ & Size of the key universe. \\
        $N$ & The number of keys in the dataset. \\
        $R$ & Maximum query range length. \\
        $\epsilon$ & Target FPR. \\
        \cmidrule{1-2}
        $h(x)$ & General hash function of an RSQF. \\
        $h_f(x)$ & The fingerprint of $x$ resulting from $h(x)$. \\
        $n$ & The number of slots of an RSQF. \\
        $F$ & The hash table of an RSQF. \\
        $\alpha$ & A RSQF's load factor. \\
        $f$ & Fingerprint size used in filter. Expressed in bits. \\
        \cmidrule{1-2}
        $r$ & Size of the mementos. Expressed in bits. \\
        $m(k)$ & The memento of a key $k$. \\
        $p(k)$ & The prefix of a key $k$. \\
        $\ell$ & Average number of keys in a non-empty partition. \\
        Memento & The $r$ least significant bits of a key. \\
        Prefix & The prefix excluding the $r$ least significant bits. \\
        Partition & A partition of the universe defined by a prefix. \\ 
        Vacant Fingerprint & A zero fingerprint used in encoding keepsake boxes. \\
        \bottomrule
    \end{tabular}
    \caption{Definitions of terms and symbols. The table is split into three
    sections presenting the terms and notation used to describe the general
    range filtering problem, the RSQF, and Memento filter, respectively.}
    \label{tab:term_and_symbol_definitions}
\end{table}

Many applications are subject to \emph{Correlated Workloads}~\cite{Rosetta,
Grafite}. The range queries of such workloads are close to the keys of the
underlying dataset but do not contain keys within them. The reason such queries
are commonplace is that users typically issue queries that are informed by the
keys in the dataset. For example, a user searching a medical database for
patients in a given age group is more likely to issue the query 40-41 years old
than 140-141. Many range filters exhibit high FPRs in the face of correlated
workloads, as they only maintain coarse-grain information about the data. The
general distance between the queries and the keys is referred to as the
\emph{Correlation Degree} of the workload. A \emph{robust} range filter
supports any workload with range queries of length at most~$R$, including
correlated workloads, without degradation of its FPR.

Goswami et al.~\cite{Goswami} prove an information-theoretic memory lower bound
of $\log_2(\frac{R}{\epsilon}) - O(1)$ bits per key for any robust range filter
supporting range queries of length at most~$R$ with an FPR of~$\epsilon$. While
several non-robust range filters have been proposed that require less memory
than this bound, they do so in exchange for a much higher FPR under correlated
workloads.

\textbf{Prior Work.}
We now describe all existing range filters in a roughly chronological order.

\textbf{ARF}~\cite{AdaptiveRangeFilter} is a range filter employing a binary
trie that adapts to the data and query distributions. However, it is superseded
by \textbf{SuRF}~\cite{SuRF}, which utilizes succinct tries to compactly encode
keys while maintaining ordering information to support range queries. This trie
contains the shortest unique prefix of each key to ensure that the final
structure does not consume too much memory, causing the number of internal
nodes of the trie to depend on the length and distribution of the keys.
Fingerprints and key suffixes can be stored in the leaves of the trie for each
key, improving support of point and range queries, respectively. 

SuRF exhibits a high FPR under correlated workloads since the shortest unique
prefixes cannot differentiate between close queries and keys. Moreover, SuRF
does not support insertions or deletes due to its succinct encoding scheme, and
its query time deteriorates with the length of keys.

\textbf{Rosetta}~\cite{Rosetta} employs a hierarchy of Bloom filters, each
storing key prefixes of a given length. The hierarchy is treated as a segment
tree~\cite{SegmentTree} and supports range filtering by checking whether all
prefixes in a specified query range are absent. Rosetta achieves a near-optimal
FPR by employing a recursive query process that corrects the
false positives of an upper-level filter with the help of lower-level filters.
During this process, Rosetta checks for $O(\log_2 R)$ sub-intervals as dictated
by the segment tree, possibly followed by additional Bloom filter lookups.
While the hierarchical structure of Rosetta makes it robust, each query entails
probing many bits chosen by hash functions, leading to many random
cache misses~\cite{REncoder, REncoder_Journal, bloomRF}.

\textbf{REncoder}~\cite{REncoder, REncoder_Journal} and
\textbf{bloomRF}~\cite{bloomRF} improve on Rosetta's speed by encoding the
prefixes of a given key in a cache-friendly manner within a single bitmap.
Specifically, REncoder breaks Rosetta's segment tree into mini-segment trees
and encodes each contiguously in multiple locations. These contiguous encodings
give the filter access to the query range decomposition with a single memory
access. In contrast, bloomRF generalizes the segment tree to have a larger
fanout based on the key length and the dataset size. It further employs a
Prefix Hashing scheme with Piecewise Monotonic Hash Functions to improve cache
locality. Such a hashing scheme positions neighboring sub-intervals of equal
length next to each other in the bitmap, giving simultaneous access to them all
using a single cache miss.

While REncoder and bloomRF improve on Rosetta’s speed, they still incur several
cache misses for queries and do not provide dynamicity or expandability. They
also forgo robustness, as encoding all the levels of the segment tree in the
same bitmap results in a uniform FPR assignment to all levels. This uniformity
significantly decreases filtering for correlated queries since only a limited
number of the lower levels of the tree can filter them out.

\textbf{Proteus}~\cite{Proteus} combines SuRF with a Bloom filter. The SuRF
instance is truncated to contain prefixes of keys up to a given length~$l_1$
and acts as a pre-filter for the Bloom filter, while the Bloom filter stores
key prefixes of a fixed length~$l_2 > l_1$. This hybrid structure exposes a
rich range filter design space with many tradeoffs. Proteus tunes the prefix
lengths $l_1$ and $l_2$ to minimize the FPR of the hybrid filter, which
improves upon both hierarchical filter designs and SuRF. Consequently, Proteus
discards prefixes longer than the longest-common prefix of the keys with the
queries, causing the structure to lose robustness against correlated queries
since it cannot differentiate between close keys and queries. Furthermore, its
tuning procedure requires both a sample query set and a static underlying
dataset. Queries are also expensive due to the many random cache misses caused
by the SuRF instance and the Bloom filter.

\textbf{SNARF}~\cite{SNARF} learns from the underlying dataset by creating a
linear spline model of the keys' cumulative distribution function. It uses this
model to map each key to a bit position in a large bit array and sets it to
one. Since this is a monotonic mapping, SNARF answers range queries by scanning
the corresponding range of bits in the bit array and returns true if it
encounters a one. SNARF saves a substantial amount of memory by compressing the
bit array using Golomb or Elias-Fano coding~\cite{Golomb, EliasFanoElias,
EliasFanoFano}. \textbf{Oasis+}~\cite{Oasis} employs SNARF's framework and
improves upon its learned mapping function by pruning large empty regions of
the key space, achieving lower FPRs. It also employs instances of
Proteus~\cite{Proteus} to answer range queries for select regions of the key
space, depending on the data distribution.

However, the FPR of both of these filters suffers under correlated workloads,
as such queries tend to always map to a one in the bit array. Moreover, these
filters assume complete knowledge of the keys to train the mapping function,
meaning they must be constructed on a static dataset. Lastly, they use a
logarithmic number of random memory accesses to query the distribution model
and use floating-point operations during the process, which slow down the
filter and cause precision issues for long keys. 

\begin{table*}
    \centering
    \scriptsize
    \def\arraystretch{1.25}
    \addtolength{\tabcolsep}{-0.2em}
    \begin{tabular}{cccccccc}
        \toprule

        \textbf{Filter} & \textbf{Construction} & \textbf{Delete} & \textbf{Range Query (-)} 
        & \textbf{Range Query (+)} & \textbf{Memory Footprint} 
        & \textbf{Robust} & \textbf{Expandable} \\

        \midrule

        SuRF & $O(N \log u)$ & - & $O(\log u)$ & $O(\log u)$ 
        & $10 + \frac{10z}{N} + m + o(1)$ & \xmark & \xmark \\

        Rosetta & $O(N\log_2({\frac{R}{\epsilon}}))$ & - & $O(\log_2(R))$ 
        & $O(\log_2({\frac{R}{\epsilon}}))$ & $1.44 \log_2({\frac{R}{\epsilon}})$ 
        & \cmark & \xmark \\
        
        REncoder * & $O(Nk)$ & - & $k$ & $k$ & $O(k + \log(\frac{1}{\epsilon}))$
        & \xmark & \xmark \\

        bloomRF * & $O(N\log(\frac{u}{N}))$ & - & $O(\log(\frac{u}{N}))$ 
        & $O(\log(\frac{u}{N}))$ & $\approx 1.2 \log_2(\frac{R}{\epsilon})$ 
        & \xmark & \xmark \\

        Proteus * & $O(N\log(\frac{u}{\epsilon}))$ & - &
        $O(\log u)$ & $O(\log(\frac{u}{\epsilon}))$ &
        $\frac{10z}{N} + 1.44 \log_2(\frac{1}{\epsilon})$
        & \xmark & \xmark \\

        SNARF * & $O(N)$ & $O(\log_2 N)$ & $O(\log_2 N)$ 
        & $O(\log_2 N)$ & $2.4 + \log_2(\frac{1}{\epsilon})$ 
        & \xmark & \xmark \\

        Oasis+ * & $O(N)$ & - & $O(\log_2 N)$ & $O(\log_2 N)$ & $\approx 2.4 +
        \log_2(\frac{1}{\epsilon})$ & \xmark & \xmark \\

        Grafite & $O(N \log_2 N)$ & - & $1-2$ & $1-2$ 
        & $2 + \log_2(\frac{R}{\epsilon}) + o(1)$ & \cmark & \xmark \\

        \cmidrule{1-8}

        \makecell[c]{Memento \\ (Cache Misses)} & $O(N)$ & $1$ & $1-2$ &
        $1-2$ & \multirow{2}{*}{$\frac{1}{\alpha} (3.125 +
        \log_2(\frac{R}{\epsilon}))$} & \multirow{2}{*}{\cmark} &
        \multirow{2}{*}{\cmark} \\
        \makecell[c]{Memento \\ (CPU)} & $O(\ell N) \approx O(N)$ &
        $O(\ell) \approx O(1)$ & $O(\log_2 \ell) \approx O(1)$ & $O(\log_2
        \ell) \approx O(1)$ &  &  & \\

        \bottomrule
    \end{tabular}
    \normalfont
    \caption{A comparison of range filters assuming an FPR of~$\epsilon$,
    maximum range query length~$R$, and $N$ keys coming from a universe of size
    $u$. For SuRF and Proteus, $z$ refers to the number of internal nodes in
    the trie, while $m$ denotes the length of the fingerprints stored at the
    leaves. For REncoder, $k$ refers to the number of hash functions, which we
    have empirically found to be $O(\log(\frac{1}{\epsilon}))$. For Memento,
    $\ell$ is a measure of the local density of the keys which is at most $R$
    and thus small. The operation costs are measured in the expected number of
    random cache misses, and the memory footprint is measured in bits per key.
    As shown, no existing method supports dynamic key sets, a robust FPR, and
    fast operations, all at the same time.}
    \label{tab:existing_method_stats}
\end{table*}

\textbf{Grafite}~\cite{Grafite} implements Goswami et al.'s design of a range
filter~\cite{Goswami}, which uses a locality-preserving hash function to map
each key to a bit in a bit array. It compresses the bit array using Elias-Fano
coding~\cite{EliasFanoElias, EliasFanoFano} and handles queries by checking the
corresponding range of bits for any set bit. The corresponding range is found
by accessing a rank-and-select structure~\cite{Poppy, FlorianRankandSelect}
built on top of the Elias-Fano encoding, followed by a binary
search~\cite{CompactPATTrees, RankandSelectDict, PartitionedEliasFano}.

Grafite is the current state of the art in terms of speed, as it requires up to
three random memory accesses to serve a query. Its hashing and coding scheme
allows it to achieve the same robustness as Rosetta while enjoying extremely
fast queries and a much better FPR vs. memory tradeoff. However, the bit array
and the hashing function do not allow for insertions, deletions, or expansions
without reconstructing the structure from scratch, as with the other Bloom
filter-inspired methods. 

\textbf{Summary.}
Table~\ref{tab:existing_method_stats} summarizes the characteristics of
existing range filters. The methods annotated with * are heuristic in nature
and do not provide strict mathematical bounds on their memory consumption.
Therefore, we provide a conservative estimate of their memory footprint based
on the experimental data from their respective papers to enable a comparison. 

Table~\ref{tab:existing_method_stats} shows that existing range filters do not
support deletes and cannot expand as more data is inserted. This makes them
inapplicable across the wide range of database applications that support range
queries over rapidly changing and/or growing data (e.g., from B-tree access in
OLTP applications to analytical queries in HTAP systems). Is it possible to
design a robust range filter that can accommodate dynamic data while also being
competitive in terms of query cost, FPR, and memory footprint?

\begin{figure}
    \centering
    \vspace{-0.75cm}
    \begin{tikzpicture}
        \def\d{1}
        \def\n{4}
        \def\nn{3}
        \def\w{0.35}
        \def\R{1}
        \def\midd{1}

        \foreach \tmp in {0, 1} {
            \draw (\midd * \tmp + \d * \n * \tmp + -0.65 * \d, \w * \R + 0.5 * \w) rectangle (\midd * \tmp + \d * \n * \tmp, \w * \R + 1.5 * \w);
            \draw[fill=gray!20] (\midd * \tmp + \d * \n * \tmp, \w * \R + \w) rectangle (\midd * \tmp + \d * \n * \tmp + \d * \n, \w * \R + \w * 2);
            \draw (\midd * \tmp + \d * \n * \tmp, \w * \R) rectangle (\midd * \tmp + \d * \n * \tmp + \d * \n, \w * \R + \w);
            \draw (\midd * \tmp + \d * \n * \tmp, 0) rectangle (\midd * \tmp + \d * \n * \tmp + \d * \n, \w * \R);
            \draw (\midd * \tmp + \d * \n * \tmp, 0) rectangle (\midd * \tmp + \d * \n * \tmp + \d * \n, \w * \R + 2 * \w);

            \foreach \x in {0, ..., \nn} {
                \draw[dashed] (\midd * \tmp + \d * \n * \tmp + \x * \d, 0) -- (\midd * \tmp + \d * \n * \tmp + \x * \d, \w * \R + \w * 2);
            }
        }
        \draw[very thick] (0, \w * \R) -- (\d * \n, \w * \R);
        \draw[very thick] (\midd + \d * \n, \w * \R) -- (\midd + 2 * \d * \n, \w * \R);

        \node[inner sep=0pt] (offset) at (-1.25 * \d, \w * \R + 1.05 * \w) {\smaller \verb|offset|};
        \node[inner sep=0pt] (occupieds) at (\midd + 2 * \n * \d + 0.9 * \d, \w * \R + 1.5 * \w) {\smaller \verb|occupieds|};
        \node[inner sep=0pt] (runends) at (\midd + 2 * \n * \d + 0.9 * \d, \w * \R + 0.5 * \w) {\smaller \verb|runends|};
        \node[inner sep=0pt] (slots) at (\midd + 2 * \n * \d + 0.9 * \d, 0.5 * \w * \R) {\smaller \verb|slots|};

        \node[inner sep=0pt] (offset_val_0) at (-0.325 * \d, \w * \R + \w) {$0$};
        \node[inner sep=0pt] (offset_val_1) at (\midd + \n * \d + -0.325 * \d, \w * \R + \w) {$2$};

        \node[inner sep=0pt] (address_0) at (0 * \d + 0.5 * \d, \w * \R + 2.5 * \w) {0};
        \node[inner sep=0pt] (address_1) at (1 * \d + 0.5 * \d, \w * \R + 2.5 * \w) {1};
        \node[inner sep=0pt] (address_2) at (2 * \d + 0.5 * \d, \w * \R + 2.5 * \w) {2};
        \node[inner sep=0pt] (address_3) at (3 * \d + 0.5 * \d, \w * \R + 2.5 * \w) {3};
        \node[inner sep=0pt] (address_4) at (\midd + 4 * \d + 0.5 * \d, \w * \R + 2.5 * \w) {4};
        \node[inner sep=0pt] (address_5) at (\midd + 5 * \d + 0.5 * \d, \w * \R + 2.5 * \w) {5};
        \node[inner sep=0pt] (address_6) at (\midd + 6 * \d + 0.5 * \d, \w * \R + 2.5 * \w) {6};
        \node[inner sep=0pt] (address_7) at (\midd + 7 * \d + 0.5 * \d, \w * \R + 2.5 * \w) {7};

        \node[inner sep=0pt] (occupieds_0) at (0 * \d + 0.5 * \d, \w * \R + 1.5 * \w) {1};
        \node[inner sep=0pt] (occupieds_1) at (1 * \d + 0.5 * \d, \w * \R + 1.5 * \w) {1};
        \node[inner sep=0pt] (occupieds_2) at (2 * \d + 0.5 * \d, \w * \R + 1.5 * \w) {0};
        \node[inner sep=0pt] (occupieds_3) at (3 * \d + 0.5 * \d, \w * \R + 1.5 * \w) {1};
        \node[inner sep=0pt] (occupieds_4) at (\midd + 4 * \d + 0.5 * \d, \w * \R + 1.5 * \w) {0};
        \node[inner sep=0pt] (occupieds_5) at (\midd + 5 * \d + 0.5 * \d, \w * \R + 1.5 * \w) {1};
        \node[inner sep=0pt] (occupieds_6) at (\midd + 6 * \d + 0.5 * \d, \w * \R + 1.5 * \w) {0};
        \node[inner sep=0pt] (occupieds_7) at (\midd + 7 * \d + 0.5 * \d, \w * \R + 1.5 * \w) {0};

        \node[inner sep=0pt] (runends_0) at (0 * \d + 0.5 * \d, \w * \R + 0.5 * \w) {0};
        \node[inner sep=0pt] (runends_1) at (1 * \d + 0.5 * \d, \w * \R + 0.5 * \w) {1};
        \node[inner sep=0pt] (runends_2) at (2 * \d + 0.5 * \d, \w * \R + 0.5 * \w) {1};
        \node[inner sep=0pt] (runends_3) at (3 * \d + 0.5 * \d, \w * \R + 0.5 * \w) {0};
        \node[inner sep=0pt] (runends_4) at (\midd + 4 * \d + 0.5 * \d, \w * \R + 0.5 * \w) {0};
        \node[inner sep=0pt] (runends_5) at (\midd + 5 * \d + 0.5 * \d, \w * \R + 0.5 * \w) {1};
        \node[inner sep=0pt] (runends_6) at (\midd + 6 * \d + 0.5 * \d, \w * \R + 0.5 * \w) {1};
        \node[inner sep=0pt] (runends_7) at (\midd + 7 * \d + 0.5 * \d, \w * \R + 0.5 * \w) {0};

        \node[inner sep=0pt] (slot_0) at (0 * \d + 0.5 * \d, 0.5 * \w * \R) {\textcolor{blue}{\textbf{0110}}};
        \node[inner sep=0pt] (slot_1) at (1 * \d + 0.5 * \d, 0.5 * \w * \R) {\textcolor{blue}{\textbf{1011}}};
        \node[inner sep=0pt] (slot_2) at (2 * \d + 0.5 * \d, 0.5 * \w * \R) {\textcolor{blue}{\textbf{1000}}};
        \node[inner sep=0pt] (slot_3) at (3 * \d + 0.5 * \d, 0.5 * \w * \R) {\textcolor{blue}{\textbf{0000}}};
        \node[inner sep=0pt] (slot_4) at (\midd + 4 * \d + 0.5 * \d, 0.5 * \w * \R) {\textcolor{blue}{\textbf{0001}}};
        \node[inner sep=0pt] (slot_5) at (\midd + 5 * \d + 0.5 * \d, 0.5 * \w * \R) {\textcolor{blue}{\textbf{0101}}};
        \node[inner sep=0pt] (slot_6) at (\midd + 6 * \d + 0.5 * \d, 0.5 * \w * \R) {\textcolor{blue}{\textbf{1111}}};

        \draw[decorate, decoration={brace, mirror, raise=1pt, amplitude=4pt}] (0 * \d, 0) -- (2 * \d, 0) node[pos=0.5, below=3pt] {\footnotesize Run};
        \draw[decorate, decoration={brace, mirror, raise=1pt, amplitude=4pt}] (2 * \d, 0) -- (3 * \d, 0) node[pos=0.5, below=3pt] {\footnotesize Run};
        \draw[decorate, decoration={brace, mirror, raise=1pt, amplitude=4pt}] (3 * \d, 0) -- (\midd + 6 * \d, 0) node[pos=0.5, below=3pt] {\footnotesize Run};
        \draw[decorate, decoration={brace, mirror, raise=1pt, amplitude=4pt}] (\midd + 6 * \d, 0) -- (\midd + 7 * \d, 0) node[pos=0.5, below=3pt] {\footnotesize Run};

        \draw[decorate, decoration={brace, raise=1pt, amplitude=4pt}] (0 * \d, \R * \w + 2.9 * \w) -- (3 * \d, \R * \w + 2.9 * \w) node[pos=0.5, above=3pt] {\footnotesize Cluster};
        \draw[decorate, decoration={brace, raise=1pt, amplitude=4pt}] (3 * \d, \R * \w + 2.9 * \w) -- (\midd + 7 * \d, \R * \w + 2.9 * \w) node[pos=0.5, above=3pt] {\footnotesize Cluster};
        \draw[dotted] (0 * \d, \R * \w + 2 * \w) -- (0 * \d, \R * \w + 2.9 * \w);
        \draw[dotted] (3 * \d, \R * \w + 2 * \w) -- (3 * \d, \R * \w + 2.9 * \w);
        \draw[dotted] (\midd + 7 * \d, \R * \w + 2 * \w) -- (\midd + 7 * \d, \R * \w + 2.9 * \w);

        \node[inner sep=0pt] (caption) at (-0.75 * \d, 0.5 * \w * \R) {\textbf{(A)}};
    \end{tikzpicture}
    \vspace{2pt}

    \begin{tikzpicture}
        \def\d{1.0}
        \def\n{4}
        \def\nn{3}
        \def\w{0.35}
        \def\R{1}
        \def\midd{1.0}

        \foreach \tmp in {0, 1} {
            \draw (\midd * \tmp + \d * \n * \tmp + -0.65 * \d, \w * \R + 0.5 * \w) rectangle (\midd * \tmp + \d * \n * \tmp, \w * \R + 1.5 * \w);
            \draw[fill=gray!20] (\midd * \tmp + \d * \n * \tmp, \w * \R + \w) rectangle (\midd * \tmp + \d * \n * \tmp + \d * \n, \w * \R + \w * 2);
            \draw (\midd * \tmp + \d * \n * \tmp, \w * \R) rectangle (\midd * \tmp + \d * \n * \tmp + \d * \n, \w * \R + \w);
            \draw (\midd * \tmp + \d * \n * \tmp, 0) rectangle (\midd * \tmp + \d * \n * \tmp + \d * \n, \w * \R);
            \draw (\midd * \tmp + \d * \n * \tmp, 0) rectangle (\midd * \tmp + \d * \n * \tmp + \d * \n, \w * \R + 2 * \w);

            \foreach \x in {0, ..., \nn} {
                \draw[dashed] (\midd * \tmp + \d * \n * \tmp + \x * \d, 0) -- (\midd * \tmp + \d * \n * \tmp + \x * \d, \w * \R + \w * 2);
            }
        }
        \draw[very thick] (0, \w * \R) -- (\d * \n, \w * \R);
        \draw[very thick] (\midd + \d * \n, \w * \R) -- (\midd + 2 * \d * \n, \w * \R);

        \node[inner sep=0pt] (spacer) at (\midd + 2 * \n * \d + 0.825 * \d, \w * \R + 0.5 * \w) { };

        \node[inner sep=0pt] (offset_val_0) at (-0.325 * \d, \w * \R + \w) {$0$};
        \node[inner sep=0pt] (offset_val_1) at (\midd + \n * \d + -0.325 * \d, \w * \R + \w) {$3$};

        \node[inner sep=0pt] (occupieds_0) at (0 * \d + 0.5 * \d, \w * \R + 1.5 * \w) {1};
        \node[inner sep=0pt] (occupieds_1) at (1 * \d + 0.5 * \d, \w * \R + 1.5 * \w) {1};
        \node[inner sep=0pt] (occupieds_2) at (2 * \d + 0.5 * \d, \w * \R + 1.5 * \w) {0};
        \node[inner sep=0pt] (occupieds_3) at (3 * \d + 0.5 * \d, \w * \R + 1.5 * \w) {1};
        \node[inner sep=0pt] (occupieds_4) at (\midd + 4 * \d + 0.5 * \d, \w * \R + 1.5 * \w) {0};
        \node[inner sep=0pt] (occupieds_5) at (\midd + 5 * \d + 0.5 * \d, \w * \R + 1.5 * \w) {1};
        \node[inner sep=0pt] (occupieds_6) at (\midd + 6 * \d + 0.5 * \d, \w * \R + 1.5 * \w) {0};
        \node[inner sep=0pt] (occupieds_7) at (\midd + 7 * \d + 0.5 * \d, \w * \R + 1.5 * \w) {0};

        \node[inner sep=0pt] (runends_0) at (0 * \d + 0.5 * \d, \w * \R + 0.5 * \w) {0};
        \node[inner sep=0pt] (runends_1) at (1 * \d + 0.5 * \d, \w * \R + 0.5 * \w) {1};
        \node[inner sep=0pt] (runends_2) at (2 * \d + 0.5 * \d, \w * \R + 0.5 * \w) {1};
        \node[inner sep=0pt] (runends_3) at (3 * \d + 0.5 * \d, \w * \R + 0.5 * \w) {0};
        \node[inner sep=0pt] (runends_4) at (\midd + 4 * \d + 0.5 * \d, \w * \R + 0.5 * \w) {0};
        \node[inner sep=0pt] (runends_5) at (\midd + 5 * \d + 0.5 * \d, \w * \R + 0.5 * \w) {0};
        \node[inner sep=0pt] (runends_6) at (\midd + 6 * \d + 0.5 * \d, \w * \R + 0.5 * \w) {1};
        \node[inner sep=0pt] (runends_7) at (\midd + 7 * \d + 0.5 * \d, \w * \R + 0.5 * \w) {1};

        \node[inner sep=0pt] (slot_0) at (0 * \d + 0.5 * \d, 0.5 * \w * \R) {\textcolor{blue}{\textbf{0110}}};
        \node[inner sep=0pt] (slot_1) at (1 * \d + 0.5 * \d, 0.5 * \w * \R) {\textcolor{blue}{\textbf{1011}}};
        \node[inner sep=0pt] (slot_2) at (2 * \d + 0.5 * \d, 0.5 * \w * \R) {\textcolor{blue}{\textbf{1000}}};
        \node[inner sep=0pt] (slot_3) at (3 * \d + 0.5 * \d, 0.5 * \w * \R) {\textcolor{blue}{\textbf{0000}}};
        \node[inner sep=0pt] (slot_4) at (\midd + 4 * \d + 0.5 * \d, 0.5 * \w * \R) {\textcolor{blue}{\textbf{0001}}};
        \node[inner sep=0pt] (slot_5) at (\midd + 5 * \d + 0.5 * \d, 0.5 * \w * \R) {\textcolor{blue}{\textbf{0101}}};
        \node[inner sep=0pt] (slot_6) at (\midd + 6 * \d + 0.5 * \d, 0.5 * \w * \R) {\textcolor{blue}{\textbf{1010}}};
        \node[inner sep=0pt] (slot_7) at (\midd + 7 * \d + 0.5 * \d, 0.5 * \w * \R) {\textcolor{blue}{\textbf{1111}}};

        \node[inner sep=0pt] (caption) at (-0.75 * \d, 0.5 * \w * \R) {\textbf{(B)}};
        \node[inner sep=0pt] (op) at (\midd + 5.75 * \d, \w * \R + 2.75 * \w) {\small Insert $x$, $h(x) = $ \textcolor{blue}{\textbf{1010}}0011};
        \draw[-stealth] (op.west) -| (3 * \d + 0.5 * \d, \w * \R + 2 * \w);
    \end{tikzpicture}
    \vspace{8pt}

    \begin{tikzpicture}
        \def\d{1.0}
        \def\n{4}
        \def\nn{3}
        \def\w{0.35}
        \def\R{1}
        \def\midd{1.0}

        \foreach \tmp in {0, 1} {
            \draw (\midd * \tmp + \d * \n * \tmp + -0.65 * \d, \w * \R + 0.5 * \w) rectangle (\midd * \tmp + \d * \n * \tmp, \w * \R + 1.5 * \w);
            \draw[fill=gray!20] (\midd * \tmp + \d * \n * \tmp, \w * \R + \w) rectangle (\midd * \tmp + \d * \n * \tmp + \d * \n, \w * \R + \w * 2);
            \draw (\midd * \tmp + \d * \n * \tmp, \w * \R) rectangle (\midd * \tmp + \d * \n * \tmp + \d * \n, \w * \R + \w);
            \draw (\midd * \tmp + \d * \n * \tmp, 0) rectangle (\midd * \tmp + \d * \n * \tmp + \d * \n, \w * \R);
            \draw (\midd * \tmp + \d * \n * \tmp, 0) rectangle (\midd * \tmp + \d * \n * \tmp + \d * \n, \w * \R + 2 * \w);

            \foreach \x in {0, ..., \nn} {
                \draw[dashed] (\midd * \tmp + \d * \n * \tmp + \x * \d, 0) -- (\midd * \tmp + \d * \n * \tmp + \x * \d, \w * \R + \w * 2);
            }
        }
        \draw[very thick] (0, \w * \R) -- (\d * \n, \w * \R);
        \draw[very thick] (\midd + \d * \n, \w * \R) -- (\midd + 2 * \d * \n, \w * \R);

        \node[inner sep=0pt] (spacer) at (\midd + 2 * \n * \d + 0.825 * \d, \w * \R + 0.5 * \w) { };

        \node[inner sep=0pt] (offset_val_0) at (-0.325 * \d, \w * \R + \w) {$0$};
        \node[inner sep=0pt] (offset_val_1) at (\midd + \n * \d + -0.325 * \d, \w * \R + \w) {$3$};

        \node[inner sep=0pt] (occupieds_0) at (0 * \d + 0.5 * \d, \w * \R + 1.5 * \w) {1};
        \node[inner sep=0pt] (occupieds_1) at (1 * \d + 0.5 * \d, \w * \R + 1.5 * \w) {1};
        \node[inner sep=0pt] (occupieds_2) at (2 * \d + 0.5 * \d, \w * \R + 1.5 * \w) {0};
        \node[inner sep=0pt] (occupieds_3) at (3 * \d + 0.5 * \d, \w * \R + 1.5 * \w) {1};
        \node[inner sep=0pt] (occupieds_4) at (\midd + 4 * \d + 0.5 * \d, \w * \R + 1.5 * \w) {0};
        \node[inner sep=0pt] (occupieds_5) at (\midd + 5 * \d + 0.5 * \d, \w * \R + 1.5 * \w) {1};
        \node[inner sep=0pt] (occupieds_6) at (\midd + 6 * \d + 0.5 * \d, \w * \R + 1.5 * \w) {0};
        \node[inner sep=0pt] (occupieds_7) at (\midd + 7 * \d + 0.5 * \d, \w * \R + 1.5 * \w) {0};

        \node[inner sep=0pt] (runends_0) at (0 * \d + 0.5 * \d, \w * \R + 0.5 * \w) {1};
        \node[inner sep=0pt] (runends_1) at (1 * \d + 0.5 * \d, \w * \R + 0.5 * \w) {1};
        \node[inner sep=0pt] (runends_2) at (2 * \d + 0.5 * \d, \w * \R + 0.5 * \w) {0};
        \node[inner sep=0pt] (runends_3) at (3 * \d + 0.5 * \d, \w * \R + 0.5 * \w) {0};
        \node[inner sep=0pt] (runends_4) at (\midd + 4 * \d + 0.5 * \d, \w * \R + 0.5 * \w) {0};
        \node[inner sep=0pt] (runends_5) at (\midd + 5 * \d + 0.5 * \d, \w * \R + 0.5 * \w) {0};
        \node[inner sep=0pt] (runends_6) at (\midd + 6 * \d + 0.5 * \d, \w * \R + 0.5 * \w) {1};
        \node[inner sep=0pt] (runends_7) at (\midd + 7 * \d + 0.5 * \d, \w * \R + 0.5 * \w) {1};

        \node[inner sep=0pt] (slot_0) at (0 * \d + 0.5 * \d, 0.5 * \w * \R) {\textcolor{blue}{\textbf{1011}}};
        \node[inner sep=0pt] (slot_1) at (1 * \d + 0.5 * \d, 0.5 * \w * \R) {\textcolor{blue}{\textbf{1000}}};
        \node[inner sep=0pt] (slot_3) at (3 * \d + 0.5 * \d, 0.5 * \w * \R) {\textcolor{blue}{\textbf{0000}}};
        \node[inner sep=0pt] (slot_4) at (\midd + 4 * \d + 0.5 * \d, 0.5 * \w * \R) {\textcolor{blue}{\textbf{0001}}};
        \node[inner sep=0pt] (slot_5) at (\midd + 5 * \d + 0.5 * \d, 0.5 * \w * \R) {\textcolor{blue}{\textbf{0101}}};
        \node[inner sep=0pt] (slot_6) at (\midd + 6 * \d + 0.5 * \d, 0.5 * \w * \R) {\textcolor{blue}{\textbf{1010}}};
        \node[inner sep=0pt] (slot_7) at (\midd + 7 * \d + 0.5 * \d, 0.5 * \w * \R) {\textcolor{blue}{\textbf{1111}}};

        \node[inner sep=0pt] (caption) at (-0.75 * \d, 0.5 * \w * \R) {\textbf{(C)}};
        \node[inner sep=0pt] (op) at (\midd + 5.75 * \d, \w * \R + 2.75 * \w) {\small Delete $y$, $h(y) = $ \textcolor{blue}{\textbf{0110}}0000};
        \draw[-stealth] (op.west) -| (0 * \d + 0.5 * \d, \w * \R + 2 * \w);
    \end{tikzpicture}
    \caption{An RSQF handles hash collisions by pushing fingerprints to the
    right using Robin Hood Hashing.}
    \label{fig:RSQF_block}
\end{figure}
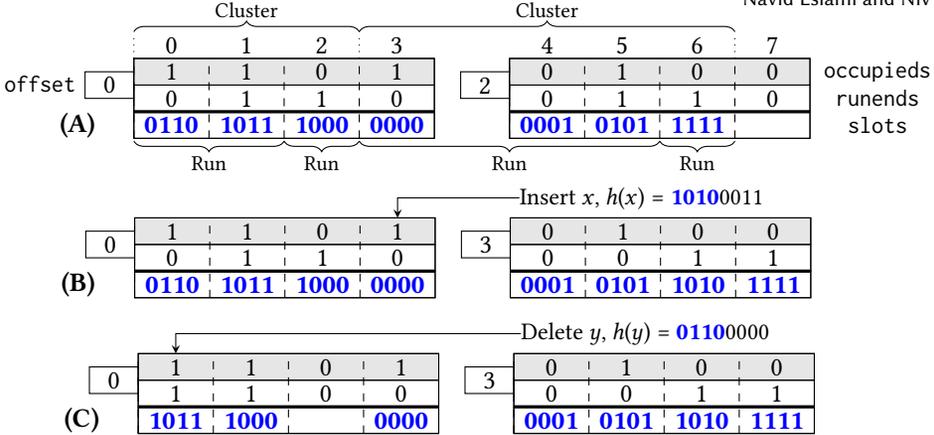

\section{Background}~\label{sec:background}
This section describes the Rank-and-Select Quotient Filter (RSQF)~\cite{GQF},
the base hash table implementation on top of which we build Memento filter.
The second section of Table~\ref{tab:term_and_symbol_definitions} lists the
symbolds we use henceforth to describe the RSQF and Memento filter. An RSQF is
a compact hash table that stores a fingerprint for each key. An RSQF's hash
table $F$ consists of $n$ slots, each able to store an $f$-bit fingerprint.
This hash table maps a key $x$ to its \emph{Canonical Slot} using the $\lceil
\log_2(n) \rceil$ least significant bits of the key's hash $h(x)$. It further
associates a fingerprint $h_f(x)$ to $x$ by taking the following $f$ bits of
the aforementioned hash value. For example, considering an RSQF with 16 slots,
a key~$x$ with a hash of $h(x) = $ \textbf{0110}0000 would have Slot 0 as its
canonical slot based on the least significant bits of its hash. It will also
have $h_f(x) = $ \textbf{0110} as its fingerprint based on the remaining bits
of $h(x)$. 

An RSQF resolves hash collisions via Robin Hood
Hashing~\cite{RobinHoodHashing}, meaning that all fingerprints mapped to a
given canonical slot are stored contiguously. To achieve this, fingerprints may
shift any other fingerprint they collide with to the right to make space for
themselves. A \emph{Run} is defined as the contiguous set of slots storing the
fingerprints corresponding to a given canonical slot. A group of contiguous
slots occupied by runs, where all but the left-most run are shifted to the
right, is known as a \emph{Cluster}. Fig.~\ref{fig:RSQF_block}-(A) shows a
populated RSQF, along with its runs and clusters. The bold blue text indicates
the fingerprints, while the black text indicates the canonical slot addresses.

\textbf{Metadata.}
An RSQF associates two metadata bits to each slot to represent where runs begin
and end: the \verb|occupieds| and \verb|runends| bits. The \verb|occupieds| bit
for slot $i$ indicates whether or not a key with canonical slot $F[i]$ was
inserted into the filter. The \verb|runends| bit indicates whether or not the
fingerprint stored at slot $F[i]$ is the last fingerprint in a run. For
example, the set \verb|occupieds| bits in Fig.~\ref{fig:RSQF_block}-(A)
indicate that Slots~0, 1, 3, and 5 are canonical slots, while the set
\verb|runends| bits indicate that the runs of these slots end at Slots~1, 2, 5,
and 6, respectively.

To improve performance, an RSQF is partitioned into blocks of 64 slots, each
augmented with two contiguous 64-bit bitmaps representing the \verb|occupieds|
and \verb|runends| bits and an 8-bit \verb|offset| field. A block's
\verb|offset| field indicates the number of fingerprints belonging to canonical
slots from previous blocks shifted into its slots or subsequent blocks. For
example, the \verb|offset| value of the second block in
Fig.~\ref{fig:RSQF_block}-(A) is equal to 2 because its first two slots contain
fingerprints with a canonical slot of 3, which is from the previous block. This
field enables an RSQF to skip over many irrelevant slots to find the run of a
key.

\textbf{Locating a Run.}
Since an RSQF shifts runs to the right, it must search for a canonical slot's
run. To this end, an RSQF leverages the fact that each one bit in the
\verb|occupieds| bitmap has a corresponding one bit in the \verb|runends|
bitmap denoting the end of its run. Therefore, it locates slot $F[i]$'s run by
finding the set \verb|runends| bit that matches $F[i]$'s \verb|occupieds| bit.
An RSQF starts this process by considering the \verb|offset| field of the
associated block. The filter skips \verb|offset| many slots to the right,
allowing it to immediately find a slot $F[j]$ containing runs from $F[i]$'s
block. Then, it uses specialized CPU instructions to apply efficient
rank-and-select operations on the \verb|occupieds| bitmap fragment of $F[i]$'s
block and the \verb|runends| bitmap fragment of $F[j]$'s block to quickly find
the matching set bits, thus locating the end of $F[i]$'s run.

For example, in Fig.~\ref{fig:RSQF_block}-(A), Slot~5's run is located by first
skipping \verb|offset|=2 slots in the block. Then, using the rank operation on
the \verb|occupieds| bitmap, an RSQF realizes that Slot~5's associated runend
bit comes first among this block's runend bits. Armed with this knowledge, an
RSQF uses the select operation on the \verb|runends| bitmap fragment, ignoring
its first \verb|offset|=2 bits, to locate the first set bit. Thus, it finds the
end of $F[i]$'s run, i.e., Slot~6.

\textbf{Queries.}
A query for a key $q$ starts at the canonical slot of $q$. If the
\verb|occupieds| flag of this slot is zero, an RSQF returns a negative since a
run for that slot does not exist. If it is one, its run is located using the
procedure described above. An RSQF then searches the discovered run for a
fingerprint equal to $h_f(q)$. If one exists, the query returns a positive and
a negative otherwise.

\textbf{Inserts.}
An RSQF inserts a key $x$ by first locating its run. It then adds the
fingerprint $h_f(x)$ to the run by shifting the subsequent runs one slot to the
right and updating the filter's metadata accordingly. This shifting procedure
may add new runs to $x$'s cluster. Fig.~\ref{fig:RSQF_block}-(B) shows the
insertion process of such a key with hash \textbf{1010}0011. Here, Slot 3's run
expands and pushes Slot 0101's run to the right. It also updates the
\verb|offset| field of the next block as it shifts slots into it.

\textbf{Deletes.}
Similarly to an insert, an RSQF deletes a key $y$ by finding its run and
removing from it a fingerprint equal to $h_f(y)$. It then shifts the subsequent
slots in the cluster to the left to keep the cluster contiguous. This may split
the cluster into smaller clusters. Fig.~\ref{fig:RSQF_block}-(C) shows an
example of a deletion to fingerprint \textbf{0110} at canonical Slot 0. Here,
the run shrinks, and the cluster splits into two small clusters of size 1 each.

\textbf{Iteration.}
RSQFs support the iteration of their fingerprints via a left-to-right scan.
Using the metadata flags, the canonical slot of each fingerprint can be
identified. An RSQF can then recover the original hash bits of a key by
concatenating its fingerprint to the address of its canonical slot. As we shall
see later, iteration serves as the cornerstone of expansion operations.

\textbf{Allocation.}
An RSQF supports a load factor of up to $\alpha = 95\%$. Beyond this load
factor, the filter's performance deteriorates rapidly since the cluster lengths
skyrocket. 

\textbf{Analysis.}
An RSQF has an FPR of $\epsilon \leq \alpha \cdot 2^{-f}$ since each canonical
slot has an average of $\alpha$ fingerprints in its run, and each of those
fingerprints matches a query with probability $2^{-f}$. The structure has an
overall memory footprint of $\frac{1}{\alpha} (2.125 + f)$ bits per key.

\begin{figure}
    \centering
    \begin{tikzpicture}
        \def\l{-3.5}
        \def\r{1.5}
        \def\d{0.75}

        \draw (\l-\d, 1.6) rectangle (-\d, 1.1);
        \node[inner sep=0pt] (p_k) at (0.5*\l-\d,1.35) {$p(k)$};
        \draw[fill=red!10] (-\d, 1.6) rectangle (\r-\d, 1.1);
        \node[inner sep=0pt] (m_k) at (0.5*\r-\d,1.35) {$m(k)$};

        \draw[decorate, decoration={brace, raise=2pt, amplitude=4pt}] (-\d, 1.6) -- (\r-\d, 1.6) node[pos=0.5, above=6pt] {$r$ bits};

        \node[inner sep=0pt] (label_A) at (-6, 1.35) {\textbf{(A)}};


        \def\R{2}
        \def\L{-4}
        \def\preflen{1}
        \def\prefcnt{6}
        \def\seplen{0.2}
        \draw[black] (\L, 0) -- (\R, 0);
        \foreach \i in {1, ..., \prefcnt} {
            \draw[black] (\L + \preflen * \i, -0.5 * \seplen) -- (\L + \preflen * \i, 0.5 * \seplen);
        }
        \draw[black, very thick] (\L, -0.5 * \seplen) -- (\L, 0.5 * \seplen);
        \draw[black, very thick] (\R, -0.5 * \seplen) -- (\R, 0.5 * \seplen);
        \node[inner sep=0pt] (u) at (\L - 0.75 * \preflen, 0) {\small Universe};

        \node[inner sep=1pt, draw, circle] () at (\L + 0.3 * \preflen, 0) {};
        \node[inner sep=1pt, draw, circle] () at (\L + 0.7 * \preflen, 0) {};
        \node[inner sep=1pt, draw, circle] () at (\L + 1.6 * \preflen, 0) {};
        \node[inner sep=1pt, draw, circle] () at (\L + 2.25 * \preflen, 0) {};
        \node[inner sep=1pt, draw, circle] () at (\L + 2.5 * \preflen, 0) {};
        \node[inner sep=1pt, draw, circle] () at (\L + 2.75 * \preflen, 0) {};
        \node[inner sep=1pt, draw, circle] () at (\L + 3.15 * \preflen, 0) {};
        \node[inner sep=1pt, draw, thick, circle, red] (k_circle) at (\L + 3.55 * \preflen, 0) {};
        \node[inner sep=1pt, draw, circle] () at (\L + 4.4 * \preflen, 0) {};
        \node[inner sep=1pt, draw, circle] () at (\L + 4.9 * \preflen, 0) {};

        \draw[black, dotted] (\L + 3 * \preflen, 0) -- (\L + 3 * \preflen, 0.4);
        \draw[black, dotted] (\L + 4 * \preflen, 0) -- (\L + 4 * \preflen, 0.4);
        \draw[decorate, decoration={brace, raise=2pt, amplitude=2pt}] 
                        (\L + 3 * \preflen, 0.4) -- (\L + 4 * \preflen, 0.4)
                        node[pos=0.5, above=3pt] {$p(k)$};

        \draw[-stealth] (0.5*\l-\d, 1.0975) -- (\L+3.15*\preflen, 0.85);

        \node[inner sep=0pt, above=0.05cm of k_circle] (k) {$k$};
        \draw[decorate, decoration={brace, mirror, raise=2pt, amplitude=2pt}] 
                        (\L + 3 * \preflen, 0) -- (\L + 3.55 * \preflen, 0)
                        node[pos=0.5, below=3pt] {$m(k)$};

        \node[inner sep=0pt] (label_B) at (-6, 0) {\textbf{(B)}};
    \end{tikzpicture}
    \caption{Each key $k$ is split into a prefix $p(k)$ and a memento $m(k)$.
    Prefixes partition the key universe and cluster the keys, while mementos
    denote the positions of keys in their partitions. Each circle in (B)
    represents a key in the key set.}
    \label{fig:key_and_universe_partitioning}
\end{figure}
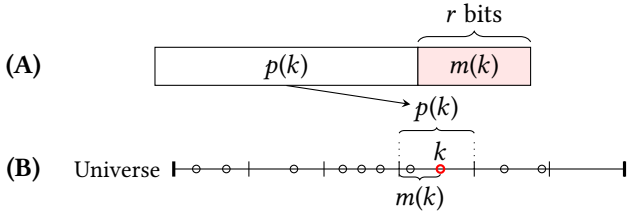

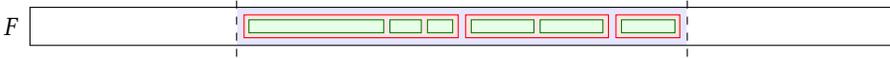
\begin{figure}
    \centering
    \begin{tikzpicture}
        \def\insetdiff{0.1}
        \def\insetmul{1.65}
        
        \draw[white,fill=blue!10] (0, 0.25) rectangle (6, 0.75);
        \draw[dashed] (0, 0.1) -- (0, 0.9);
        \draw[dashed] (6, 0.1) -- (6, 0.9);

        \draw[red,fill=red!10] (0 + \insetdiff, 0.25 + \insetdiff) rectangle (3 - 0.5 * \insetdiff, 0.75 - \insetdiff);
        \draw[red,fill=red!10] (3 + 0.5 * \insetdiff, 0.25 + \insetdiff) rectangle (5 - 0.5 * \insetdiff, 0.75 - \insetdiff);
        \draw[red,fill=red!10] (5 + 0.5 * \insetdiff, 0.25 + \insetdiff) rectangle (6 - \insetdiff, 0.75 - \insetdiff);

        \draw[green!50!black,fill=green!10] (0 + \insetmul * \insetdiff, 0.25 + \insetmul * \insetdiff) rectangle (2 - 0.25 * \insetmul * \insetdiff, 0.75 - \insetmul * \insetdiff);
        \draw[green!50!black,fill=green!10] (2 + 0.25 * \insetmul * \insetdiff, 0.25 + \insetmul * \insetdiff) rectangle (2.5 - 0.25 * \insetmul * \insetdiff, 0.75 - \insetmul * \insetdiff);
        \draw[green!50!black,fill=green!10] (2.5 + 0.25 * \insetmul * \insetdiff, 0.25 + \insetmul * \insetdiff) rectangle (3 - 0.75 * \insetmul * \insetdiff, 0.75 - \insetmul * \insetdiff);

        \draw[green!50!black,fill=green!10] (3 + 0.75 * \insetmul * \insetdiff, 0.25 + \insetmul * \insetdiff) rectangle (4 - 0.25 * \insetmul * \insetdiff, 0.75 - \insetmul * \insetdiff);
        \draw[green!50!black,fill=green!10] (4 + 0.25 * \insetmul * \insetdiff, 0.25 + \insetmul * \insetdiff) rectangle (5 - 0.75 * \insetmul * \insetdiff, 0.75 - \insetmul * \insetdiff);

        \draw[green!50!black,fill=green!10] (5 + 0.75 * \insetmul * \insetdiff, 0.25 + \insetmul * \insetdiff) rectangle (6 - \insetmul * \insetdiff, 0.75 - \insetmul * \insetdiff);

        \draw (-2.75, 0.25) rectangle (8.75, 0.75);
        \node[inner sep=0pt] (F) at (-3, 0.5) {$F$};
    \end{tikzpicture}
    \caption{Clusters are comprised of runs, where all but the first are
    shifted to the right. A run contains one or more keepsake boxes. The single
    cluster in the illustration has three runs, and the runs contain three,
    two, and one keepsake box, respectively.}
    \label{fig:nestings}
\end{figure}

\section{Memento Filter}~\label{sec:memento_filter}
We introduce Memento filter, the first range filter to simultaneously provide a
robust FPR, fast inserts, queries, and deletes, and the ability to efficiently
expand and contract. Conceptually, Memento filter partitions the key space into
equally sized partitions. For each partition with at least one key, it
contiguously stores a fingerprint along with fixed-length suffixes of all keys
in that partition. We refer to these suffixes as mementos, and we refer to a
fingerprint along with its collection of mementos as a keepsake box. Memento
filter processes a range query by visiting all intersecting partitions with a
matching fingerprint and checking for overlapping mementos in their keepsake
boxes. Memento filter guarantees a desired FPR for fixed-length keys and range
queries of length up to $R$. We show how to extend it to support
variable-length keys at the expense of robustness and arbitrary range queries
at the expense of FPR and speed.

We build Memento filter on top of an RSQF as its use of Robin Hood Hashing
allows for storing variable-length keepsake boxes. At the same time, such
variable-length keepsake boxes may elongate the runs and clusters of the filter
and potentially damage performance. To counteract this, we show how to
succinctly encode keepsake boxes and how to traverse them efficiently, leading
to 1 and 2 random cache misses for point and range queries, respectively. In
Section~\ref{sec:theoretical_analysis}, we show analytically that Memento
Filter achieves the same performance as an RSQF by leveraging this succinct
encoding. Table~\ref{tab:term_and_symbol_definitions} provides a list of
symbols and definitions used in this section. 

\textbf{Prefixes and Mementos.}
Memento filter splits each key $k$ into a \emph{prefix} $p(k)$ and a
\emph{memento} $m(k)$. A memento is the $r = \lceil \log_2 R \rceil$ least
significant bits of $k$, where $R$ is the maximum range query length that the
filter must support. A prefix is the maximal prefix of $k$ not containing the
$r$ least significant bits. Fig.~\ref{fig:key_and_universe_partitioning}-(A)
provides an example of this split. Based on these prefixes, Memento filter
partitions the \emph{key universe} into partitions of length $2^r \geq R$, as
depicted in Fig.~\ref{fig:key_and_universe_partitioning}-(B). A partition
contains keys with the same prefix. Given this partitioning, a key $k$'s
memento $m(k)$ represents $k$'s position in the partition defined by its prefix
$p(k)$. Notice that, as in the design of all range filters (except SuRF), we
assume the keys to be fixed-length strings.

Since the maximum query range size $R$ is typically small in comparison to the
universe size $u$, a key's prefix tends to be much longer than its memento,
e.g., a 56-bit prefix vs. an 8-bit memento for a 64-bit key. Furthermore, since
the partition size $2^r$ is at least as large as the query length $R$, any
range query intersects with at most two consecutive partitions of the key
universe. Thus, Memento filter splits a query range into at most two
sub-ranges, each subsumed by a partition. It then answers the query by checking
for the inclusion of any mementos in these sub-ranges.

Crucially, the resulting range query performance is workload-agnostic. The
intuition is that mementos encode information about each key's least
significant bits. Hence, any query's end points can be reliably checked for
overlap with the mementos within the matching partitions to ascertain
membership with a false positive rate that depends only on the fingerprint
length.

Memento filter uses the prefix of the keys to insert them into an RSQF. That
is, it computes $h(p(k))$ and $h_f(p(k))$ to map a key $k$ to a canonical slot
and to derive a fingerprint, where $h(\cdot)$ and $h_f(\cdot)$ are hash
functions defined in Section~\ref{sec:background} in the context of an RSQF.
Memento filter stores the mementos and the fingerprint of the keys in each
partition consecutively and succinctly to maximize cache locality and space
efficiency.

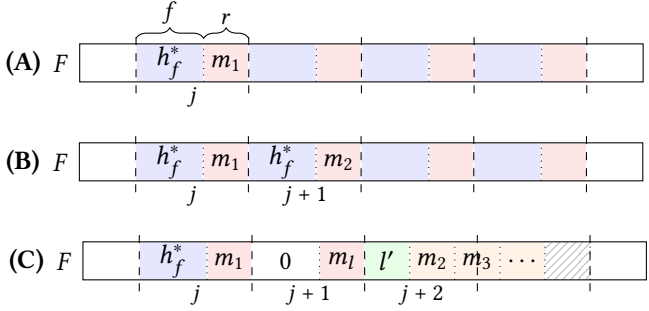
\begin{figure}
    \centering
    \begin{tikzpicture}
        \foreach \x in {0, ..., 3} {
            \draw[white, fill=blue!10] (1.5 * \x, 0.25) rectangle (1.5 * \x + 0.9, 0.75);
            \draw[white, fill=red!10] (1.5 * \x + 0.9, 0.25) rectangle (1.5 * \x + 1.5, 0.75);
            \draw[dotted] (1.5 * \x + 0.9, 0.25) -- (1.5 * \x + 0.9, 0.75);
            \draw[dashed] (1.5 * \x, 0.1) -- (1.5 * \x, 0.9);
        }
        \draw[dashed] (1.5 * 4, 0.1) -- (1.5 * 4, 0.9);
        \draw (-0.75, 0.25) rectangle (6.75, 0.75);
        \draw[decorate, decoration={brace, raise=2pt, amplitude=4pt}] (0, 0.75) -- (0.9, 0.75) node[pos=0.5, above=4pt] {\small $f$};
        \draw[decorate, decoration={brace, raise=2pt, amplitude=4pt}] (0.9, 0.75) -- (1.5, 0.75) node[pos=0.5, above=4pt] {\small $r$};

        \node[inner sep=0pt] (label) at (-1.5, 0.5) {\textbf{(A)}};
        \node[inner sep=0pt] (F) at (-1, 0.5) {$F$};
        \node[inner sep=0pt] (canonical_slot) at (0.75, 0.05) {\small $j$};
        \node[inner sep=0pt] (prefix_set_fingerprint) at (0.45, 0.5) {$h_f^*$};
        \node[inner sep=0pt] (memento) at (1.2, 0.5) {$m_1$};
    \end{tikzpicture}
    \vspace{8pt}

    \begin{tikzpicture}
        \foreach \x in {0, ..., 3} {
            \draw[white, fill=blue!10] (1.5 * \x, 0.25) rectangle (1.5 * \x + 0.9, 0.75);
            \draw[white, fill=red!10] (1.5 * \x + 0.9, 0.25) rectangle (1.5 * \x + 1.5, 0.75);
            \draw[dotted] (1.5 * \x + 0.9, 0.25) -- (1.5 * \x + 0.9, 0.75);
            \draw[dashed] (1.5 * \x, 0.1) -- (1.5 * \x, 0.9);
        }
        \draw[dashed] (1.5 * 4, 0.1) -- (1.5 * 4, 0.9);
        \draw (-0.75, 0.25) rectangle (6.75, 0.75);

        \node[inner sep=0pt] (label) at (-1.5, 0.5) {\textbf{(B)}};
        \node[inner sep=0pt] (F) at (-1, 0.5) {$F$};
        \node[inner sep=0pt] (canonical_slot) at (0.75, 0.05) {\small $j$};
        \node[inner sep=0pt] (prefix_set_fingerprint) at (0.45, 0.5) {$h_f^*$};
        \node[inner sep=0pt] (memento) at (1.2, 0.5) {$m_1$};
        \node[inner sep=0pt] (plus_1_slot) at (2.25, 0.05) {\small $j + 1$};
        \node[inner sep=0pt] (void_fingerprint) at (1.95, 0.5) {$h_f^*$};
        \node[inner sep=0pt] (memento_2) at (2.7, 0.5) {$m_2$};
    \end{tikzpicture}
    \vspace{8pt}

    \begin{tikzpicture}
        \foreach \x in {0, ..., 1} {
            \ifthenelse{\x=1} {
                \draw[white] (1.5 * \x, 0.25) rectangle (1.5 * \x + 0.9, 0.75);
            }{
                \draw[white, fill=blue!10] (1.5 * \x, 0.25) rectangle (1.5 * \x + 0.9, 0.75);
            }
            \draw[white, fill=red!10] (1.5 * \x + 0.9, 0.25) rectangle (1.5 * \x + 1.5, 0.75);
            \draw[dotted] (1.5 * \x + 0.9, 0.25) -- (1.5 * \x + 0.9, 0.75);
        }
        \foreach \x in {0, ..., 3} {
            \ifthenelse{\x=0} {
                \draw[white, fill=green!10] (3 + 0.6 * \x, 0.25) rectangle (3 + 0.6 * \x + 0.6, 0.75);
            }{
                \draw[white, fill=orange!10] (3 + 0.6 * \x, 0.25) rectangle (3 + 0.6 * \x + 0.6, 0.75);
            }
        }
        \draw[white,pattern=north east lines,pattern color=gray!50] (5.4, 0.25) rectangle (6, 0.75);
        \foreach \x in {0, ..., 4} {
            \draw[dotted] (3 + 0.6 * \x + 0.6, 0.25) -- (3 + 0.6 * \x + 0.6, 0.75);
        }
        \foreach \x in {0, ..., 3} {
            \draw[dashed] (1.5 * \x, 0.1) -- (1.5 * \x, 0.9);
        }
        \draw[dashed] (1.5 * 4, 0.1) -- (1.5 * 4, 0.9);
        \draw (-0.75, 0.25) rectangle (6.75, 0.75);

        \node[inner sep=0pt] (label) at (-1.5, 0.5) {\textbf{(C)}};
        \node[inner sep=0pt] (F) at (-1, 0.5) {$F$};
        \node[inner sep=0pt] (canonical_slot) at (0.75, 0.05) {\small $j$};
        \node[inner sep=0pt] (prefix_set_fingerprint) at (0.45, 0.5) {$h_f^*$};
        \node[inner sep=0pt] (memento) at (1.2, 0.5) {$m_1$};
        \node[inner sep=0pt] (plus_1_slot) at (2.25, 0.05) {\small $j + 1$};
        \node[inner sep=0pt] (void_fingerprint) at (1.95, 0.5) {$0$};
        \node[inner sep=0pt] (memento_2) at (2.7, 0.5) {$m_l$}; \node[inner
        sep=0pt] (plus_1_slot) at (3.75, 0.05) {\small $j + 2$};

        \node[inner sep=0pt] (l) at (3.3, 0.5) {$l'$}; \node[inner sep=0pt]
        (memento_2) at (3.9, 0.5) {$m_2$}; \node[inner sep=0pt] (memento_3) at
        (4.5, 0.5) {$m_3$};
        \node[inner sep=0pt] (memento_4) at (5.125, 0.5) {$\dots$};
    \end{tikzpicture}
    \caption{Each keepsake box is encoded using one of three cases, depending
    on how many mementos it contains.}
    \label{fig:keepsake_box_encoding}
\end{figure}

\textbf{Slot Structure.}
Memento filter allocates slots with a width of $f + r$ bits in its underlying
RSQF, as shown in Fig.~\ref{fig:keepsake_box_encoding}-(A). This allows it to
store one fingerprint and one memento in each slot. We omit the
\verb|occupieds|, \verb|runends|, and \verb|offset| fields of the RSQF in
Fig.~\ref{fig:keepsake_box_encoding}-(A) to highlight new design elements built
on top.

\textbf{Keepsake Boxes.}
Since Memento filter hashes the prefix of each key to map it to a canonical
slot and to generate a fingerprint, keys from different partitions may map to
the same run and may even share a fingerprint due to hash collisions. We use
the term \emph{keepsake box} to refer to the union of the keys/mementos with a
shared fingerprint in a run. Note that the keys within a keepsake box may come
from different partitions due to hash collisions. Fig.~\ref{fig:nestings} shows
how there can be multiple keepsake boxes within a run and several runs within a
cluster. Clusters and runs are delimited using the RSQF's metadata fields as
shown in Section~\ref{sec:background}. We now focus on encoding and delimiting
keepsake boxes in a run.

\textbf{Encoding of a Keepsake Box.}
Because keepsake boxes are variable-length, their encoding must represent how
long they are to facilitate unambiguous decoding. The simplest solution is to
store a counter for each keepsake box that denotes how many mementos it
contains. This approach, however, would use excessive space for metadata when
the keepsake boxes contain few mementos. 

To overcome this challenge, our filter enforces the following invariant:
\emph{the keepsake boxes of a run must be stored in non-decreasing order of
their fingerprints.} We show how this allows for delimiting keepsake boxes
within a run without using additional metadata. Furthermore, to save space and
optimize cache behavior, Memento filter minimizes the number of shared
fingerprints stored for a keepsake box. 

Consider a keepsake box with canonical slot $F[i]$, fingerprint $h_f^*$, and a
list of $l$ associated mementos $m_1 \leq \dots \leq~m_l$. Assume that this
keepsake box's fingerprint is stored in $F[j]$, where $j \geq i$ due to hash
collisions. It is encoded as follows:

\textit{Case~(1) $l = 1$}: The only memento $m_1$ is stored with the keepsake
box's fingerprint in the same slot of $F[j]$, as shown in
Fig.~\ref{fig:keepsake_box_encoding}-(A).

\textit{Case~(2) $l = 2$}: A fingerprint-memento pair is stored for each key,
as depicted in Fig.~\ref{fig:keepsake_box_encoding}-(B). The smaller memento
$m_1$ is stored with the keepsake box's fingerprint in $F[j]$, while $m_2$ is
stored with a copy of the keepsake box's fingerprint in $F[j + 1]$. 

\textit{Case~(3) $l > 2$}: The smallest memento $m_1$ is stored alongside the
keepsake box's fingerprint in $F[j]$ while the largest memento $m_l \geq m_1$
is stored with a zero \emph{vacant fingerprint} in $F[j + 1]$. The decrease in
the fingerprint values in the run created by the vacant fingerprint acts as an
escape sequence, signaling that the keepsake box has more than two members.
The smallest and largest mementos of the keepsake box stored in $F[j]$ and $F[j
+ 1]$ allow for quickly ruling out the existence of a key range without
traversing the entire keepsake box.

The rest of the mementos are encoded as a sorted list in the subsequent slots,
as shown in Fig.~\ref{fig:keepsake_box_encoding}-(C). This sorted list is
encoded by first writing down its length $l' = l - 2$ using $r$ bits on
average. This length parameter is followed up by the mementos, stored
compactly, disregarding alignment. Due to this misalignment, the last slot of
this memento list may have unused space. This unused space corresponds to the
hatched area in Fig.~\ref{fig:keepsake_box_encoding}-(C). 

Based on the above encoding scheme, Memento filter encodes each key in at most
$f + r$ bits. With more dataset skew, Memento filter forgoes storing
fingerprints for the keys, and thus the memory it uses for each key approaches
$r$ bits. A memento usually comprises one byte to be able to answer short range
queries efficiently. At the same time, the fingerprint size tends to be at
least one byte to achieve an FPR in the range of 1-10\%. This implies that each
slot can house at least two mementos. Hence, all keepsake box encodings consume
at most one slot per key. 

A minor caveat is that keepsake boxes with a fingerprint of zero cannot utilize
the vacant fingerprint as an escape sequence in Case~(3), as it does not create
a decreasing order for them. In such a scenario, Memento filter encodes the
keepsake box entirely using Case~(1). That is, each memento will have its own
fingerprint. We will see that there are no zero fingerprints in the context of
an expandable Memento filter, and this corner case will naturally disappear.

Since keepsake boxes are ordered according to their fingerprints, an increase
in the fingerprint values signals the start of a new keepsake box, which
delimits keepsake boxes in Cases~(1) and (2). In Case~(3), Memento filter
delimits keepsake boxes using the length field $l'$. Note that keepsake box
encodings are considered part of their run. Therefore, all of their slots,
except for the final slot of the run, have zero \verb|runends| bits.

\textbf{Variable-Length Counter Encoding.}
A keepsake box encoding in Case~(3) uses a length field $l'$ to record the
number of mementos in the keepsake box. This length is usually smaller than
$2^r - 1$, i.e., the maximum length that $r$ bits can represent. However, $l'$
can also exceed this threshold in the unlikely event of fingerprint collisions
of densely populated partitions. To keep the encoding small when the count is
small while still supporting the rare event of large counts, we employ a
variable-length encoding for $l'$. To this end, Memento filter reserves the
value $2^r - 1$, i.e., $r$ one bits, as a special value for $l'$ and generates
an encoding in $r$-bit chunks. 

This encoding is specifically designed to keep the common case of small counts
as performant, space-efficient, and general as possible, which is not achieved
by traditional encoding schemes. That is, any $l' < 2^r - 1$ is encoded in
binary using $r$ bits. For larger lengths, the core idea is to represent $l'$
in base-$(2^r - 1)$. Memento filter achieves this by first writing $\lfloor
\log_{2^r - 1} l' \rfloor$ copies of the value $2^r - 1$, similarly to unary
coding. $c$ of these values signals that there are $c + 1$ digits in $l'$'s
base-$(2^r - 1)$ representation. This ``unary code" is followed up with the
base-$(2^r - 1)$ representation of $l'$. Since $l'$ in base-$(2^r - 1)$ cannot
have any digit equal to $2^r - 1$, the unary code preceding it is unambiguous
and is used to recover $c$.

For example, if $r = 5$, the number $l' = 30$ is represented as a single
$r$-bit value of $\langle 30 \rangle$. However, given $l' = 31$, its base-$(2^r
- 1) = 31$ representation is $\langle 1, 0 \rangle$, which no longer has a
single digit. Therefore, $l'$ is encoded to $\langle 31, 1, 0 \rangle$. This
encoding has $c = 1$ values of $2^r - 1 = 31$, implying that the base-31
representation of $l'$ has $c + 1 = 2$ digits. The code is then finished by
appending $l'$'s base-$(2^r - 1)$ representation $\langle 1, 0 \rangle$. As
$l'$ grows, this base-31 encoding is updated accordingly, e.g., $l'=32$ is
encoded to $\langle 31, 1, 1 \rangle$.

Note that Memento filter still uses at most one slot per memento with this
encoding scheme. The reason is that $l'$ is encoded using more than a single
$r$-bit chunk only when it is very large. In this case, the succinct encoding
of the long memento list compensates for the extra space required by the length
encoding.

\textbf{Skipping Keepsake Boxes.}
Memento filter skips over large keepsake boxes with mismatching fingerprints to
dramatically improve lookup speed. Since large keepsake boxes are encoded using
Case~(3), Memento filter uses the list length $l'$ to infer and skip the
appropriate number of slots to access the next keepsake box.

\textbf{Insertions.}
Memento filter inserts a key $x$ by following the semantics of its underlying
RSQF. It first finds $x$'s canonical slot using $h(p(x))$ and searches for its
run. If it finds no such run, it creates one and encodes $x$'s keepsake box in
it. Otherwise, Memento filter iterates over the run and looks for a keepsake
box associated with $x$, skipping the contents of irrelevant keepsake boxes
along the way. If there is no keepsake box with a fingerprint matching $x$'s
partition, Memento filter creates a keepsake box and positions it in the run
such that the fingerprints maintain a non-decreasing order. Otherwise, $m(x)$
is added to the matching keepsake box, updating the encoding according to the
various cases shown in Fig.~\ref{fig:keepsake_box_encoding}. The insertion
procedure may shift the filter's slots to the right, potentially merging
several clusters. 

\textbf{Deletions.}
Memento filter deletes a key $y$ by first finding its keepsake box in its run.
It then locates and removes some memento equal to $m(y)$ in the keepsake box.
Similarly to how an RSQF handles deletes, Memento filter may have to shift
several slots to the left, potentially splitting their cluster into smaller
clusters.

\begin{figure}
    \centering
    \begin{tikzpicture}
        \def\R{0.5}
        \def\L{-0.5}
        \def\preflen{2}
        \draw[black] (\R, 0) -- (\R + 2 * \preflen, 0);
        \draw[black] (\R, -0.1) -- (\R, 0.1);
        \draw[black] (\R + \preflen, -0.1) -- (\R + \preflen, 0.1);
        \draw[black] (\R + 2 * \preflen, -0.1) -- (\R + 2 * \preflen, 0.1);

        \node[inner sep=0pt] (p_l) at (\R + 0.5 * \preflen, 0.3) {$p(q_l)$};
        \node[inner sep=0pt] (p_r) at (\R + 1.5 * \preflen, 0.3) {$p(q_r)$};

        \draw[thick] (\R + 0.5 * \preflen, -0.7) -- (\R + 1.5 * \preflen, -0.7);
        \draw[thick] (\R + 0.5 * \preflen, -0.8) -- (\R + 0.5 * \preflen, -0.6);
        \draw[thick] (\R + 1.5 * \preflen, -0.8) -- (\R + 1.5 * \preflen, -0.6);

        \node[inner sep=0pt] (q_2) at (\R + \preflen, -0.5) {$q$};

        \node[inner sep=1pt, draw, circle, very thick, red] () at (\R + 1.115 * \preflen, 0) {};
        \node[inner sep=1pt, draw, circle] () at (\R + 1.3 * \preflen, 0) {};
        \node[inner sep=1pt, draw, circle] () at (\R + 1.66 * \preflen, 0) {};

        \node[inner sep=1pt, draw, circle, very thick, red] () at (\R + 0.7 * \preflen, 0) {};
        \node[inner sep=1pt, draw, circle] () at (\R + 0.11 * \preflen, 0) {};

        \node[inner sep=0pt] (subfigure_b) at (\R + \preflen, -1.1) {\textbf{(B)}};


        \draw[black] (\L - 0.5 * \preflen, 0) -- (\L - 1.5 * \preflen, 0);
        \draw[black] (\L - 0.5 * \preflen, -0.1) -- (\L - 0.5 * \preflen, 0.1);
        \draw[black] (\L - 1.5 * \preflen, -0.1) -- (\L - 1.5 * \preflen, 0.1);

        \node[inner sep=0pt] (p_l) at (\L - \preflen, 0.3) {$p^*$};

        \draw[thick] (\L - 0.7 * \preflen, -0.7) -- (\L - 1.1 * \preflen, -0.7);
        \draw[thick] (\L - 0.7 * \preflen, -0.8) -- (\L - 0.7 * \preflen, -0.6);
        \draw[thick] (\L - 1.1 * \preflen, -0.8) -- (\L - 1.1 * \preflen, -0.6);

        \node[inner sep=0pt] (q_1) at (\L - 0.9 * \preflen, -0.5) {$q$};

        \node[inner sep=1pt, draw, circle] () at (\L - 1.115 * \preflen, 0) {};
        \node[inner sep=1pt, draw, circle] () at (\L - 1.3 * \preflen, 0) {};
        \node[inner sep=1pt, draw, circle, very thick, red] () at (\L - 0.8 * \preflen, 0) {};

        \node[inner sep=0pt] (subfigure_a) at (\L - \preflen, -1.1) {\textbf{(A)}};
    \end{tikzpicture}
    \caption{A range query $q=[q_l, q_r]$ spans (A)~one or (B)~two partitions.
    The keys of a partition are represented as circles, and the checked lower
    bound, largest, and smallest keys are highlighted in bold.}
    \label{fig:range_query_cases}
\end{figure}
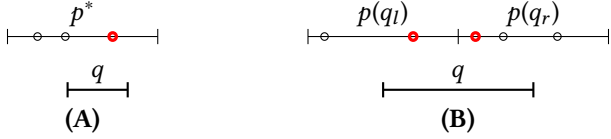

\textbf{Range Queries.}
As with all range filters, we consider range queries with a maximum length
of~$R$. Any such range query $q=[q_l, q_r]$ may intersect with at most two
partitions of the key universe since the partitions are of length $2^r$ and
$q_r - q_l + 1 \leq R \leq 2^r$. Thus, it must be the case that $p(q_r) -
p(q_l) \leq 1$. That is, the prefixes of the end-points of the query can differ
by at most one. Given this observation, Memento filter processes a range query
as follows:

\textit{If $p(q_l) = p(q_r) = p^*$}:
In this case, the query intersects with a single partition, i.e., the set of
keys with a prefix equal to $p^*$, as depicted in
Fig.~\ref{fig:range_query_cases}-(A). Memento filter first checks if the
canonical slot of the partition $p^*$ is occupied using the \verb|occupieds|
bitmap. If it is not, the query results in a negative, as no keepsake box
exists for $p^*$. If it is, Memento filter searches for a keepsake box with a
fingerprint matching $h_f(p^*)$. It returns a negative if such a keepsake box
does not exist. If one does exist, $m(q_r)$'s \emph{lower bound}, i.e., the
largest memento in the keepsake box that is less than or equal to $m(q_r)$, is
calculated using binary search and is checked for inclusion in the range
$[m(q_l), m(q_r)]$. If this range includes the lower bound, the query results
in a positive, since it signifies that a potential key of the key set is in the
query range. Otherwise, the query returns a negative, as no key is in the
range. In Fig.~\ref{fig:range_query_cases}-(A), the bold circle is the lower
bounding memento of $m(q_r)$. Here, the example query results in a positive
since the lower bounding memento lies in the range $[m(q_l), m(q_r)]$, and
because the relative ordering of the mementos in a partition matches the
ordering of the keys.

\begin{figure}
    \centering
    \begin{tikzpicture}
        \def\d{1.4}
        \def\n{8}
        \def\nn{7}
        \def\w{0.35}
        \def\R{1}

        \draw (0, \w * \R) rectangle (\d * \n, \w * \R + 2 * \w);

        \foreach \x in {0, ..., \nn} {
            \draw[fill=gray!20] (\x * \d, \w * \R + \w) rectangle (\x * \d + 0.5 * \d, \w * \R + \w * 2);
            \draw (\x * \d + 0.5 * \d, \w * \R + \w) rectangle (\x * \d + \d, \w * \R + \w * 2);
            \draw[dashed] (\x * \d, \w * \R) -- (\x * \d, \w * \R + \w * 2);
        }

        \node[inner sep=0pt,align=center] (occupiedsrunends) at (\n * \d + 1.0 * \d, \w * \R + 1.5 * \w) {\small \verb|occupieds|/\verb|runends|};
        \node[inner sep=0pt] (slots) at (\n * \d + 1.0 * \d, \w * \R + 0.5 * \w) {\small \verb|slots|};

        \begin{scope}[name prefix=address_]
            \node[inner sep=2pt] (0) at (0 * \d + 0.5 * \d, \w * \R + 2.5 * \w) {0};
            \node[inner sep=2pt] (1) at (1 * \d + 0.5 * \d, \w * \R + 2.5 * \w) {1};
            \node[inner sep=2pt] (2) at (2 * \d + 0.5 * \d, \w * \R + 2.5 * \w) {2};
            \node[inner sep=2pt] (3) at (3 * \d + 0.5 * \d, \w * \R + 2.5 * \w) {3};
            \node[inner sep=2pt] (4) at (4 * \d + 0.5 * \d, \w * \R + 2.5 * \w) {4};
            \node[inner sep=2pt] (5) at (5 * \d + 0.5 * \d, \w * \R + 2.5 * \w) {5};
            \node[inner sep=2pt] (6) at (6 * \d + 0.5 * \d, \w * \R + 2.5 * \w) {6};
            \node[inner sep=2pt] (7) at (7 * \d + 0.5 * \d, \w * \R + 2.5 * \w) {7};
        \end{scope}

        \begin{scope}[name prefix=occupieds_]
            \node[inner sep=0pt] (0) at (0 * \d + 0.25 * \d, \w * \R + 1.5 * \w) {1};
            \node[inner sep=0pt] (1) at (1 * \d + 0.25 * \d, \w * \R + 1.5 * \w) {1};
            \node[inner sep=0pt] (2) at (2 * \d + 0.25 * \d, \w * \R + 1.5 * \w) {0};
            \node[inner sep=0pt] (3) at (3 * \d + 0.25 * \d, \w * \R + 1.5 * \w) {0};
            \node[inner sep=0pt] (4) at (4 * \d + 0.25 * \d, \w * \R + 1.5 * \w) {1};
            \node[inner sep=0pt] (5) at (5 * \d + 0.25 * \d, \w * \R + 1.5 * \w) {1};
            \node[inner sep=0pt] (6) at (6 * \d + 0.25 * \d, \w * \R + 1.5 * \w) {0};
            \node[inner sep=0pt] (7) at (7 * \d + 0.25 * \d, \w * \R + 1.5 * \w) {0};
        \end{scope}

        \begin{scope}[name prefix=runends_]
            \node[inner sep=0pt] (0) at (0 * \d + 0.75 * \d, \w * \R + 1.5 * \w) {0};
            \node[inner sep=0pt] (1) at (1 * \d + 0.75 * \d, \w * \R + 1.5 * \w) {0};
            \node[inner sep=0pt] (2) at (2 * \d + 0.75 * \d, \w * \R + 1.5 * \w) {1};
            \node[inner sep=0pt] (3) at (3 * \d + 0.75 * \d, \w * \R + 1.5 * \w) {1};
            \node[inner sep=0pt] (4) at (4 * \d + 0.75 * \d, \w * \R + 1.5 * \w) {1};
            \node[inner sep=0pt] (5) at (5 * \d + 0.75 * \d, \w * \R + 1.5 * \w) {0};
            \node[inner sep=0pt] (6) at (6 * \d + 0.75 * \d, \w * \R + 1.5 * \w) {0};
            \node[inner sep=0pt] (7) at (7 * \d + 0.75 * \d, \w * \R + 1.5 * \w) {1};
        \end{scope}

        \begin{scope}[name prefix=slot_]
            \node[inner sep=0pt] (0) at (0 * \d + 0.5 * \d, 1.5 * \w * \R) {\textcolor{blue}{\textbf{1011}}$\;\;$\textcolor{red}{\textit{01}}};
            \node[inner sep=0pt] (1) at (1 * \d + 0.5 * \d, 1.5 * \w * \R) {0000$\;\;$\textcolor{red}{\textit{11}}};
            \node[inner sep=0pt] (2) at (2 * \d + 0.5 * \d, 1.5 * \w * \R) {\textcolor[rgb]{0,0.5,0}{01}$\;$\textcolor{red}{\textit{10}}$\;$\textcolor{gray}{00}};
            \node[inner sep=0pt] (3) at (3 * \d + 0.5 * \d, 1.5 * \w * \R) {\textcolor{blue}{\textbf{1111}}$\;\;$\textcolor{red}{\textit{01}}};
            \node[inner sep=0pt] (4) at (4 * \d + 0.5 * \d, 1.5 * \w * \R) {\textcolor{blue}{\textbf{0000}}$\;\;$\textcolor{red}{\textit{10}}};
            \node[inner sep=0pt] (5) at (5 * \d + 0.5 * \d, 1.5 * \w * \R) {\textcolor{blue}{\textbf{1101}}$\;\;$\textcolor{red}{\textit{00}}};
            \node[inner sep=0pt] (6) at (6 * \d + 0.5 * \d, 1.5 * \w * \R) {0000$\;\;$\textcolor{red}{\textit{11}}};
            \node[inner sep=0pt] (7) at (7 * \d + 0.5 * \d, 1.5 * \w * \R) {\textcolor[rgb]{0,0.5,0}{01}$\;$\textcolor{red}{\textit{01}}$\;$\textcolor{gray}{00}};
        \end{scope}

        \def\L{3}
        \def\R{11}
        \def\dy{2.5}
        \def\prefcnt{8}
        \def\preflen{1}
        \def\quantumcnt{32}
        \def\quantumlen{0.25}
        \def\seplen{0.2}
        \draw[black] (\L, \dy) -- (\R, \dy);
        \foreach \i in {1, ..., \prefcnt} {
            \draw[black] (\L + \preflen * \i, \dy - 0.5 * \seplen) -- (\L + \preflen * \i, \dy + 0.5 * \seplen);
        }
        \foreach \i in {1, ..., \quantumcnt} {
            \draw[black,dotted] (\L + \quantumlen * \i, \dy - 0.5 * \seplen) -- (\L + \quantumlen * \i, \dy + 0.5 * \seplen);
        }
        \draw[black, very thick] (\L, \dy - 0.5 * \seplen) -- (\L, \dy + 0.5 * \seplen);
        \draw[black, very thick] (\R, \dy - 0.5 * \seplen) -- (\R, \dy + 0.5 * \seplen);
        \node[inner sep=0pt] (u) at (\L - 0.75 * \preflen, \dy) {\small Universe};

        \node[inner sep=1pt, draw, circle] at (\L + 0 * \preflen + 1 * \quantumlen, \dy) {};

        \node[inner sep=1pt, draw, circle] at (\L + 3 * \preflen + 1 * \quantumlen, \dy) {};
        \node[inner sep=1pt, draw, circle] at (\L + 3 * \preflen + 2 * \quantumlen, \dy) {};
        \node[inner sep=1pt, draw, circle] at (\L + 3 * \preflen + 3 * \quantumlen, \dy) {};

        \node[inner sep=1pt, draw, circle] at (\L + 5 * \preflen + 0 * \quantumlen, \dy) {};
        \node[inner sep=1pt, draw, circle] at (\L + 5 * \preflen + 1 * \quantumlen, \dy) {};
        \node[inner sep=1pt, draw, circle] at (\L + 5 * \preflen + 3 * \quantumlen, \dy) {};

        \node[inner sep=1pt, draw, circle] at (\L + 6 * \preflen + 2 * \quantumlen, \dy) {};

        \draw[decorate, decoration={brace, mirror, raise=1pt, amplitude=4pt}] (\L + 0 * \preflen, \dy - 0.5 * \seplen) -- (\L + 1 * \preflen, \dy - 0.5 * \seplen) node[pos=0.5, below=3pt, inner sep=1pt] (mapping_1) {};
        \draw[-stealth] (mapping_1.south) -- (address_1.north);
        \draw[decorate, decoration={brace, mirror, raise=1pt, amplitude=4pt}] (\L + 3 * \preflen, \dy - 0.5 * \seplen) -- (\L + 4 * \preflen, \dy - 0.5 * \seplen) node[pos=0.5, below=3pt, inner sep=1pt] (mapping_2) {};
        \draw[-stealth] (mapping_2.south) -- (address_0.north);
        \draw[decorate, decoration={brace, mirror, raise=1pt, amplitude=4pt}] (\L + 5 * \preflen, \dy - 0.5 * \seplen) -- (\L + 6 * \preflen, \dy - 0.5 * \seplen) node[pos=0.5, below=3pt, inner sep=1pt] (mapping_3) {};
        \draw[-stealth] (mapping_3.south) -- (address_5.north);
        \draw[decorate, decoration={brace, mirror, raise=1pt, amplitude=4pt}] (\L + 6 * \preflen, \dy - 0.5 * \seplen) -- (\L + 7 * \preflen, \dy - 0.5 * \seplen) node[pos=0.5, below=3pt, inner sep=1pt] (mapping_4) {};
        \draw[-stealth] (mapping_4.south) -- (address_4.north);

        \draw[very thick] (0 * \d, \w) rectangle (3 * \d, 2 * \w);
        \draw[very thick] (3 * \d, \w) rectangle (4 * \d, 2 * \w);
        \draw[very thick] (4 * \d, \w) rectangle (5 * \d, 2 * \w);
        \draw[very thick] (5 * \d, \w) rectangle (8 * \d, 2 * \w);

        \node[inner sep=2pt] (point_query) at (\L + 0 * \preflen + 3 * \quantumlen, \dy + 3 * \seplen) {$q_3=3$};
        \draw[-stealth] (point_query.south) -- (\L + 0 * \preflen + 3 * \quantumlen, \dy);
        \node[inner sep=4pt] (range_query_single) at (\L + 3 * \preflen + 2.5 * \quantumlen, \dy + 3 * \seplen) {$q_1=[14,15]$};
        \draw[thick] ($(range_query_single.south) + (-0.125,0.0)$) -- ($(range_query_single.south) + (0.125,0.0)$);
        \draw[thick] ($(range_query_single.south) + (-0.125,-0.075)$) -- ($(range_query_single.south) + (-0.125,0.075)$);
        \draw[thick] ($(range_query_single.south) + (0.125,-0.075)$) -- ($(range_query_single.south) + (0.125,0.075)$);
        \node[inner sep=5pt] (range_query_double) at (\L + 6 * \preflen + 1 * \quantumlen, \dy + 3 * \seplen) {$q_2=[23,27]$};
        \draw[thick] ($(range_query_double.south) + (-0.5,0.0)$) -- ($(range_query_double.south) + (0.5,0.0)$);
        \draw[thick] ($(range_query_double.south) + (-0.5,-0.075)$) -- ($(range_query_double.south) + (-0.5,0.075)$);
        \draw[thick] ($(range_query_double.south) + (0.5,-0.075)$) -- ($(range_query_double.south) + (0.5,0.075)$);

        \node[inner sep=1pt] (fingerprint) at (0.2 * \d, 0) {\textcolor{blue}{\small \textbf{Fingerprint}}};
        \draw[-stealth,blue] ($(slot_0.south) + (-0.25,0.0)$) -- (fingerprint.north);
        \node[inner sep=1pt] (memento) at (1.275 * \d, 0.03) {\textcolor{red}{\small \textit{Memento}}};
        \draw[-stealth,red] ($(slot_0.south) + (0.45,0.0)$) -- (memento.150);
        \node[inner sep=1pt,align=center] (vacant_fingerprint) at (2.3 * \d, -0.05) {\small Vacant \\[-4pt] \small Fingerprint};
        \draw[-stealth] ($(slot_1.south) + (-0.25,0.0)$) -- ($(vacant_fingerprint.165) + (0.25,0.0)$);
        \node[inner sep=1pt] (list_len) at (3.45 * \d, 0.02) {\small \textcolor[rgb]{0,0.5,0}{List Length}};
        \draw[-stealth,green!50!black] ($(slot_2.south) + (-0.35,0.0)$) -- (list_len.170);
        \node[inner sep=1pt] (unused) at (4.7 * \d, 0.02) {\textcolor{gray}{\small Unused Space}};
        \draw[-stealth,gray] ($(slot_2.south) + (0.5,0.0)$) -- (unused.170);

    \end{tikzpicture}
    \caption{Memento filter services range queries by finding the keepsake
    boxes corresponding to the overlapping partitions in the key universe and
    searching for mementos that fall into the query range. Here, the filter's
    fingerprint and memento sizes are $f=4$ and $r=2$, respectively. Each
    universe partition is of length 4, and the keys are denoted by circles.
    Black arrows represent the canonical slot each partition maps to, and runs
    are delineated using thick lines. All runs begin in their respective
    canonical slots except for the first partition's run, which is pushed from
    Slot~1 to Slot~3 due to Robin Hood Hashing.}
    \label{fig:query_example}
    \vspace{-0.5cm}
\end{figure}
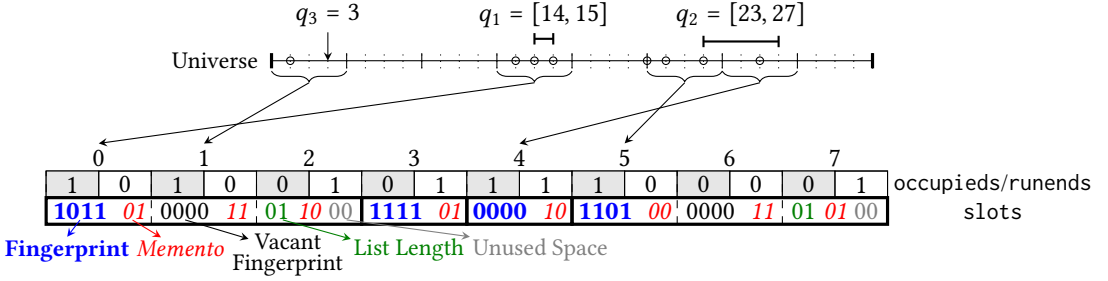

Query $q_1=[14,15]$ in Fig.~\ref{fig:query_example} is an example of a range
query intersecting a single partition. Here, mementos are $r=2$ bits long, and
both end-points have the same prefix $p^*=3$. Memento filter processes $q_1$ by
hashing the shared prefix to derive a canonical slot address and a fingerprint,
which in this example happen to be 0 and 1011, respectively. It then locates
the partition's run in Slot~0 and finds a matching keepsake box in the same
slot. Finally, it searches the keepsake box for the lower bound of the right
end-point's memento $m(q_{2_r})=3$. Since Slot~1 is in the same run and has a
vacant fingerprint, it signals that the keepsake box is encoded using Case~(3).
Thus, Memento filter searches the list of mementos stored in Slot~2 (which has
a single memento) while also accounting for the mementos in Slots~0 and 1,
resulting in a lower bound of 2. As this lower bound equals the memento of the
left end-point $m(q_{2_l})=2$, Memento filter returns a positive.

\textit{If $p(q_l) + 1 = p(q_r)$}:
This case implies that the query range has two relevant partitions: one
covering the left end-point $q_l$ and another covering the right end-point
$q_r$. Fig.~\ref{fig:range_query_cases}-(B) illustrates this case. Processing
this type of query amounts to checking whether the largest key in the partition
of $p(q_l)$ or the smallest key in the partition of $p(q_r)$ is in the query
range. One can observe from Fig.~\ref{fig:range_query_cases}-(B) that knowledge
of these points is enough to answer this range query. This is equivalent to
checking whether the largest and smallest mementos of the keys in these
partitions are contained in the sub-ranges of the range query, as defined by
the universe partitioning.

Hence, Memento filter processes this query by first locating the keepsake box
of $p(q_l)$, if it exists. If it does, its largest memento $m'$ is checked for
inclusion in $[m(q_l), 2^r - 1]$. If $m'$ is included in the range, Memento
filter returns a positive. If $m'$ is not in the range or a keepsake box for
$p(q_l)$ does not exist, the keepsake box of $p(q_r)$ is located. Memento
filter then fetches the smallest memento $m''$ from the keepsake box and checks
for its inclusion in $[0, m(q_r)]$. If $m''$ is in this range, the query
results in a positive. Otherwise, it results in a negative.
Fig.~\ref{fig:range_query_cases}-(B) shows a positive query, as $m'$ and $m''$
are in their sub-ranges, implying that the red keys are in $q$.

A crucial property of the keepsake encoding scheme is that the largest and
smallest mementos are always either stored in the same slot as the fingerprint
or in its subsequent slot. Therefore, queries find these extrema in a
cache-friendly manner without searching.

Query $q_2=[23,27]$ in Fig.~\ref{fig:query_example} is an example of a range
query intersecting two partitions. Following the above procedure, Memento
filter hashes the prefix of the left end-point $p(q_{2_l})=5$ to derive the
canonical slot address 4 and fingerprint 1101. It then finds the corresponding
run in Slot~5 and finds a matching keepsake box in the same slot. As the
keepsake box is encoded using Case (3), Memento filter determines the largest
memento in it by reading the memento stored alongside the vacant fingerprint
without searching the list of mementos starting in Slot~7. Finally, as this
largest memento, i.e., 3, equals the memento of the left end-point
$m(q_{3_l})=3$, Memento filter returns a positive. Notice how Memento filter
skipped locating the run of the right end-point, as the result of the first
probe made the query a positive.

\textbf{Longer Range Queries.} 
Memento filter also supports longer range queries by checking more partitions
in exchange for higher FPR and query times. Here, Memento filter only checks
the largest and smallest mementos for the first and last partitions, similarly
to the above discussion. Moreover, it only needs to check whether a matching
fingerprint exists for the intermediate partitions.

\textbf{Point Queries.}
Memento filter processes a point query for a key $q$ by finding its
corresponding keepsake box. The query results in a negative if there is no such
keepsake box. Otherwise, Memento filter uses binary search to find a memento
equal to $m(q)$ in the keepsake box. If it finds one, it returns true. The
absence of $m(q)$ in the keepsake box implies that $q$ was not in the key set,
and the filter thus returns a negative.

Query $q_3=3$ in Fig.~\ref{fig:query_example} is an example of a point query.
Here, Memento filter hashes the prefix to compute the partitions' canonical
slot address $h(p(q_3))=1$ and the fingerprint $h_f(p(q_3))=1111$. As the
relevant canonical slot is occupied, Memento filter locates its corresponding
run, which is in Slot~3. It iterates over the keepsake boxes and sees only one
with a matching fingerprint of 1111. It then searches for a memento equal to
$m(q_3)=3$ in the keepsake box and returns a negative result since it finds
none.

Notice that the insertion, deletion, and query operations described above are
general in the sense that they can be applied in any intermixed order, as no
assumptions are made regarding the previously applied operations.

\textbf{Bulk Loading.} 
Memento filter supports bulk loading by first sorting the keys to be inserted
in increasing order of their slot addresses, fingerprints, and mementos,
respectively. This ordering enables Memento filter to encode all runs and
keepsake boxes via a single left-to-right pass of the filter, maximizing cache
efficiency. While this is an $O(N \log N)$ algorithm, it performs better than
inserting the keys one by one in $O(\ell N) \approx O(N)$ time (as proven in
Section~\ref{sec:theoretical_analysis}), since it incurs sequential rather than
random memory accesses, leveraging the hardware prefetcher.

\textbf{Concurrency.} 
As Memento filter is built on top of the RSQF, it can reuse its concurrency
mechanisms~\cite{GQF}. More concretely, Memento filter's underlying RSQF is
partitioned into regions of 4096 slots, each with a spinlock. A thread
performing an operation locks the region its key's prefix hashes to and its
subsequent region before modifying the filter. Locking two consecutive regions
allows for thread-safe shifting of slots.

In many cases, the workload is not heavily skewed, and threads will typically
map to and lock different regions, thanks to the uniformity of hashing.
However, if the dataset is heavily skewed, many threads may want to modify the
same keepsake boxes, causing lock contention. Alleviating this contention is a
promising direction for future work. One potential approach may be buffering
blocked inserts and deletes in small, per-thread Memento filters before dumping
them into the main filter, similar to \cite{GQF}.

\textbf{Supporting Variable-Length Keys.} 
Memento filter assumes its keys to be fixed-length strings. Many applications
that use range filters operate on numerical data, which are fixed-length binary
strings. Memento filter applies to these cases as-is. One can also convert
variable-length keys into $l^*$-bit strings by zero-padding short and
truncating long keys, where $l^*$ is the smallest length such that the keys are
distinguishable using $l^*$-bit prefixes. This method strives to support the
longest range queries but forgoes robustness due to the truncated suffixes. One
can preserve robustness by increasing $l^*$ to keep the bits differentiating
the keys and queries. 

\section{Expandability}
\textbf{InfiniFilter.}
InfiniFilter is an expandable Quotient Filter (a simpler but less efficient
version of an RSQF). In InfiniFilter, each slot contains a unary ``\emph{age
counter}" of the form \underline{$0 \dots 01$} that signals how many expansions
ago a key was inserted, along with a fingerprint~\cite{InfiniFilter}. During an
expansion, a bit from each fingerprint is transferred to its canonical slot's
address, incrementing the age counter and allowing InfiniFilter to uniformly
map it to a larger filter with the same slot width as before. We call the
concatenation result of an age counter with its fingerprint a ``\emph{fluid
fingerprint}."

\begin{figure}
    \centering
    \begin{tikzpicture}
        \def\d{1.2}
        \def\n{4}
        \def\nn{3}
        \def\w{0.35}
        \def\R{1}

        \draw (0, \w * \R) rectangle (\d * \n, \w * \R + \w);
        \draw (0, 0) rectangle (\d * \n, \w * \R);
        \draw (0, 0) rectangle (\d * \n, \w * \R + 2 * \w);

        \foreach \x in {0, ..., \nn} {
            \draw[fill=gray!20] (\x * \d, \w * \R + \w) rectangle (\x * \d + 0.5 * \d, \w * \R + \w * 2);
            \draw (\x * \d + 0.5 * \d, \w * \R + \w) rectangle (\x * \d + \d, \w * \R + \w * 2);
            \draw[dashed] (\x * \d, 0) -- (\x * \d, \w * \R + \w * 2);
        }

        \node[inner sep=0pt] (occupiedsrunends) at (\n * \d + 1.25 * \d, \w * \R + 1.6 * \w) {\smaller \verb|occupieds/runends|};
        \node[inner sep=0pt] (age_counter) at (\n * \d + 1.25 * \d, \w * \R + 0.5 * \w) {\smaller \textcolor{blue}{\underline{Age Counters}}};
        \node[inner sep=0pt] (slots) at (\n * \d + 1.25 * \d, 0.5 * \w * \R - 0.1 * \w) {\smaller \verb|rest of slots|};

        \node[inner sep=0pt] (address_0) at (0 * \d + 0.5 * \d, \w * \R + 2.5 * \w) {00};
        \node[inner sep=0pt] (address_1) at (1 * \d + 0.5 * \d, \w * \R + 2.5 * \w) {01};
        \node[inner sep=0pt] (address_2) at (2 * \d + 0.5 * \d, \w * \R + 2.5 * \w) {10};
        \node[inner sep=0pt] (address_3) at (3 * \d + 0.5 * \d, \w * \R + 2.5 * \w) {11};

        \node[inner sep=0pt] (occupieds_0) at (0 * \d + 0.25 * \d, \w * \R + 1.5 * \w) {1};
        \node[inner sep=0pt] (occupieds_1) at (1 * \d + 0.25 * \d, \w * \R + 1.5 * \w) {0};
        \node[inner sep=0pt] (occupieds_2) at (2 * \d + 0.25 * \d, \w * \R + 1.5 * \w) {1};
        \node[inner sep=0pt] (occupieds_3) at (3 * \d + 0.25 * \d, \w * \R + 1.5 * \w) {0};

        \node[inner sep=0pt] (runends_0) at (0 * \d + 0.75 * \d, \w * \R + 1.5 * \w) {0};
        \node[inner sep=0pt] (runends_1) at (1 * \d + 0.75 * \d, \w * \R + 1.5 * \w) {1};
        \node[inner sep=0pt] (runends_2) at (2 * \d + 0.75 * \d, \w * \R + 1.5 * \w) {0};
        \node[inner sep=0pt] (runends_3) at (3 * \d + 0.75 * \d, \w * \R + 1.5 * \w) {1};

        \node[inner sep=0pt] (age_counter_0) at (0 * \d + 0.5 * \d, 1.48 * \w * \R) {\small \textcolor{blue}{\underline{001}}$\;$\verb|    |};
        \node[inner sep=0pt] (age_counter_1) at (1 * \d + 0.5 * \d, 1.48 * \w * \R) {\small \textcolor{blue}{\underline{01}}$\;$\verb|     |};
        \node[inner sep=0pt] (age_counter_2) at (2 * \d + 0.5 * \d, 1.48 * \w * \R) {\small \textcolor{blue}{\underline{1}}$\;$\verb|      |};

        \node[inner sep=0pt] (slot_0) at (0 * \d + 0.5 * \d, 0.5 * \w * \R) {\small \verb|   |\textcolor{blue}{\textbf{1}}$\;$\textcolor{red}{\textit{00}}};
        \node[inner sep=0pt] (slot_1) at (1 * \d + 0.5 * \d, 0.5 * \w * \R) {\small \verb|  |\textcolor{blue}{\textbf{00}}$\;$\textcolor{red}{\textit{01}}};
        \node[inner sep=0pt] (slot_2) at (2 * \d + 0.5 * \d, 0.5 * \w * \R) {\small \verb| |\textcolor{blue}{\textbf{101}}$\;$\textcolor{red}{\textit{10}}};
        \node[inner sep=0pt] (slot_3) at (3 * \d + 0.5 * \d, 0.5 * \w * \R) {\small 0000$\;$\textcolor{red}{\textit{11}}};

        \draw[very thick] (0 * \d, 0) rectangle (2 * \d, 2 * \w);
        \draw[very thick] (2 * \d, 0) rectangle (4 * \d, 2 * \w);

        \node[inner sep=0pt] (caption) at (-2.5 * \d, 0.5 * \w * \R + \w) {\textbf{(A)}};
        \node[inner sep=0pt] (fingerprint) at (-1.15 * \d, 1.20 * \w * \R) {\small \textcolor{blue}{\textbf{Fingerprint}}};
        \node[inner sep=0pt] (memento) at (-1.15 * \d, 0.25 * \w * \R) {\small \textcolor{red}{\textit{Memento}}};
        \node[inner sep=1pt,align=center] (vacant_fingerprint) at (\n * \d + 1.25 * \d, -1.4 * \w * \R) {\small Vacant Fluid \\[-4pt] \small Fingerprint};

        \draw[stealth-, blue] plot[hobby] coordinates {(-0.41 * \d, 1.20 * \w * \R) (-0.1 * \d, 1.0 * \w * \R) (0.3 * \d, 0.5 * \w * \R) (0.5 * \d, 0.5 * \w * \R)};
        \draw[stealth-, red] plot[hobby] coordinates {(-0.6 * \d, 0.25 * \w * \R) (-0.5 * \d, 0.2 * \w * \R) (0.4 * \d, -0.2 * \w * \R)  (0.8 * \d, 0.2 * \w * \R)};
        \draw[-stealth] plot[hobby] coordinates {(\n * \d - 0.65 * \d, 0.2 * \w * \R) (\n * \d - 0.25 * \d, -0.5 * \w * \R) (\n * \d + 0.25 * \d, -1.1 * \w * \R) (vacant_fingerprint.175)};


        \def\d{1.2}
        \def\dx{-2 * \d}
        \def\dy{-2.3}
        \def\n{8}
        \def\nn{7}
        \def\w{0.35}
        \def\R{1}

        \draw (\dx, \dy + \w * \R) rectangle (\dx + \d * \n, \dy + \w * \R + \w);
        \draw (\dx, \dy) rectangle (\dx + \d * \n, \dy + \w * \R);
        \draw (\dx, \dy) rectangle (\dx + \d * \n, \dy + \w * \R + 2 * \w);

        \foreach \x in {0, ..., \nn} {
            \draw[fill=gray!20] (\dx + \x * \d, \dy + \w * \R + \w) rectangle (\dx + \x * \d + 0.5 * \d, \dy + \w * \R + \w * 2);
            \draw (\dx + \x * \d + 0.5 * \d, \dy + \w * \R + \w) rectangle (\dx + \x * \d + \d, \dy + \w * \R + \w * 2);
            \draw[dashed] (\dx + \x * \d, \dy) -- (\dx + \x * \d, \dy +  \w * \R + \w * 2);
        }

        \node (address_0) at (\dx + 0 * \d + 0.5 * \d, \dy + \w * \R + 2.5 * \w) {000};
        \node (address_1) at (\dx + 1 * \d + 0.5 * \d, \dy + \w * \R + 2.5 * \w) {001};
        \node (address_2) at (\dx + 2 * \d + 0.5 * \d, \dy + \w * \R + 2.5 * \w) {010};
        \node (address_3) at (\dx + 3 * \d + 0.5 * \d, \dy + \w * \R + 2.5 * \w) {011};
        \node (address_4) at (\dx + 4 * \d + 0.5 * \d, \dy + \w * \R + 2.5 * \w) {100};
        \node (address_5) at (\dx + 5 * \d + 0.5 * \d, \dy + \w * \R + 2.5 * \w) {101};
        \node (address_6) at (\dx + 6 * \d + 0.5 * \d, \dy + \w * \R + 2.5 * \w) {110};
        \node (address_7) at (\dx + 7 * \d + 0.5 * \d, \dy + \w * \R + 2.5 * \w) {111};

        \node[inner sep=0pt] (occupieds_0) at (\dx + 0 * \d + 0.25 * \d, \dy + \w * \R + 1.5 * \w) {1};
        \node[inner sep=0pt] (occupieds_1) at (\dx + 1 * \d + 0.25 * \d, \dy + \w * \R + 1.5 * \w) {0};
        \node[inner sep=0pt] (occupieds_2) at (\dx + 2 * \d + 0.25 * \d, \dy + \w * \R + 1.5 * \w) {0};
        \node[inner sep=0pt] (occupieds_3) at (\dx + 3 * \d + 0.25 * \d, \dy + \w * \R + 1.5 * \w) {0};
        \node[inner sep=0pt] (occupieds_4) at (\dx + 4 * \d + 0.25 * \d, \dy + \w * \R + 1.5 * \w) {1};
        \node[inner sep=0pt] (occupieds_5) at (\dx + 5 * \d + 0.25 * \d, \dy + \w * \R + 1.5 * \w) {0};
        \node[inner sep=0pt] (occupieds_6) at (\dx + 6 * \d + 0.25 * \d, \dy + \w * \R + 1.5 * \w) {1};
        \node[inner sep=0pt] (occupieds_7) at (\dx + 7 * \d + 0.25 * \d, \dy + \w * \R + 1.5 * \w) {0};

        \node[inner sep=0pt] (runends_0) at (\dx + 0 * \d + 0.75 * \d, \dy + \w * \R + 1.5 * \w) {0};
        \node[inner sep=0pt] (runends_1) at (\dx + 1 * \d + 0.75 * \d, \dy + \w * \R + 1.5 * \w) {1};
        \node[inner sep=0pt] (runends_2) at (\dx + 2 * \d + 0.75 * \d, \dy + \w * \R + 1.5 * \w) {0};
        \node[inner sep=0pt] (runends_3) at (\dx + 3 * \d + 0.75 * \d, \dy + \w * \R + 1.5 * \w) {0};
        \node[inner sep=0pt] (runends_4) at (\dx + 4 * \d + 0.75 * \d, \dy + \w * \R + 1.5 * \w) {1};
        \node[inner sep=0pt] (runends_5) at (\dx + 5 * \d + 0.75 * \d, \dy + \w * \R + 1.5 * \w) {0};
        \node[inner sep=0pt] (runends_6) at (\dx + 6 * \d + 0.75 * \d, \dy + \w * \R + 1.5 * \w) {0};
        \node[inner sep=0pt] (runends_7) at (\dx + 7 * \d + 0.75 * \d, \dy + \w * \R + 1.5 * \w) {1};

        \node[inner sep=0pt] (slot_0) at (\dx + 0 * \d + 0.5 * \d, \dy + 1.48 * \w * \R) {\small \textcolor{blue}{\underline{001}}$\;$\verb|    |};
        \node[inner sep=0pt] (slot_1) at (\dx + 1 * \d + 0.5 * \d, \dy + 1.48 * \w * \R) {\small \textcolor{blue}{\underline{1}}$\;$\verb|      |};
        \node[inner sep=0pt] (slot_4) at (\dx + 4 * \d + 0.5 * \d, \dy + 1.48 * \w * \R) {\small \textcolor{blue}{\underline{0001}}$\;$\verb|   |};
        \node[inner sep=0pt] (slot_6) at (\dx + 6 * \d + 0.5 * \d, \dy + 1.48 * \w * \R) {\small \textcolor{blue}{\underline{01}}$\;$\verb|     |};

        \node[inner sep=0pt] (slot_0) at (\dx + 0 * \d + 0.5 * \d, \dy + 0.5 * \w * \R) {\small \verb|   |\textcolor{blue}{\textbf{0}}$\;$\textcolor{red}{\textit{01}}};
        \node[inner sep=0pt] (slot_1) at (\dx + 1 * \d + 0.5 * \d, \dy + 0.5 * \w * \R) {\small \verb| |\textcolor{blue}{\textbf{110}}$\;$\textcolor{red}{\textit{00}}};
        \node[inner sep=0pt] (slot_4) at (\dx + 4 * \d + 0.5 * \d, \dy + 0.5 * \w * \R) {\small \verb|    |\textcolor{blue}{}$\;$\textcolor{red}{\textit{00}}};
        \node[inner sep=0pt] (slot_6) at (\dx + 6 * \d + 0.5 * \d, \dy + 0.5 * \w * \R) {\small \verb|  |\textcolor{blue}{\textbf{10}}$\;$\textcolor{red}{\textit{10}}};
        \node[inner sep=0pt] (slot_7) at (\dx + 7 * \d + 0.5 * \d, \dy + 0.5 * \w * \R) {\small 0000$\;$\textcolor{red}{\textit{11}}};

        \draw[very thick] (\dx + 0 * \d, \dy) rectangle (\dx + 2 * \d, \dy + 2 * \w);
        \draw[very thick] (\dx + 4 * \d, \dy) rectangle (\dx + 5 * \d, \dy + 2 * \w);
        \draw[very thick] (\dx + 6 * \d, \dy) rectangle (\dx + 8 * \d, \dy + 2 * \w);

        \node[inner sep=0pt] (caption) at (\dx + -0.5 * \d, \dy + 0.5 * \w * \R + \w) {\textbf{(B)}};
        \node[inner sep=0pt] (op) at (\dx + 4.85 * \d, \dy - 1.0 * \w) 
                {Insert $x$, $h(p(x)) = $ \textcolor{blue}{\textbf{110}}000, \textcolor{red}{$m(x)=$ \textit{0}}};

        
        \draw[-stealth] (0.5 * \d, 0) -- (address_4.north);
        \draw[-stealth] (1.5 * \d, 0) -- (address_0.north);
        \draw[decorate, decoration={brace, mirror, raise=1pt, amplitude=4pt}] (2 * \d, 0) -- (4 * \d, 0) node[pos=0.5, below=3pt, inner sep=1pt] (keepsake_box_proxy) {};
        \draw[-stealth] (keepsake_box_proxy.south) -- (address_6.north);

        \draw[-stealth] (op.west) -| (\dx + 0.5 * \d, \dy);
    \end{tikzpicture}
    \caption{An expandable Memento filter employs fluid fingerprints comprised
    of an age counter and a fingerprint. It uses them to remap keepsake boxes
    during expansions. Runs are delineated using thick lines, and the slot
    addresses are shown in binary to illustrate how repurposing a fingerprint
    bit as an address bit affects the keepsake box mappings.}
    \label{fig:expandable_Memento_filter_example}
\end{figure}
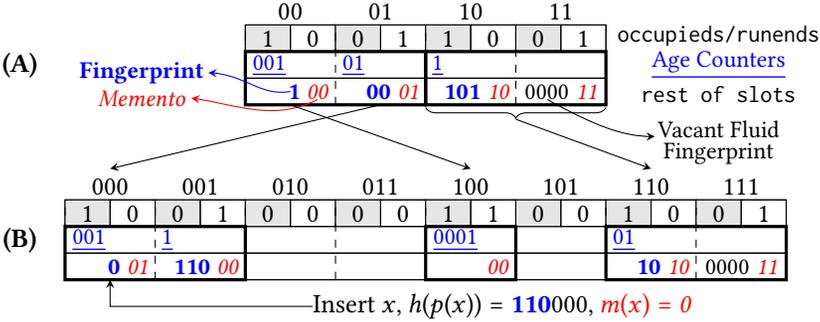

\textbf{Expandable Memento Filter.}
An expandable Memento filter stores fluid fingerprints instead of standard
fingerprints. Fig.~\ref{fig:expandable_Memento_filter_example}-(A) shows an
example of an expandable Memento filter with a fluid fingerprint and memento
length of $f = 4$ and $r = 2$. Slot 00 stores a single-bit fingerprint
\textbf{1}, along with an age counter of \underline{001}, signaling that the
key was inserted into the filter two expansions ago. 

To delimit and encode keepsake boxes within a run as described in
Section~\ref{sec:memento_filter}, they are stored in increasing order of their
fluid fingerprints. For example, in
Fig.~\ref{fig:expandable_Memento_filter_example}-(A), the run at Slot 00 stores
two keepsake boxes of length one and puts the one with a fluid fingerprint of
\underline{001}\textbf{1} before the other with a fluid fingerprint of
\underline{01}\textbf{00}.

Since the age counters always have a set bit, fluid fingerprints are never
zero. Thus, zero ``\emph{vacant fluid fingerprints}" can always act as escape
sequences for encoding keepsake boxes, eliminating the ``zero fingerprint"
corner case presented in Section~\ref{sec:memento_filter} for Case~(3).

We now describe how Memento filter uses fluid fingerprints to implement its
various operations.

\textbf{Expansions and Contractions.}
Memento filter expands by allocating a filter two times its size with the same
slot width of $f + r$ bits. Recall that the canonical slot address and the
fingerprint of each keepsake box are parts of the same original hash $h(p)$,
where $p$ is the prefix of the keepsake box's keys. With this in mind, Memento
filter iterates over the old filter, reconstructs the original hash of each
keepsake box by concatenating its fingerprint to its canonical slot address,
and inserts it into the new filter. The new canonical slot address of a
keepsake box is the $\log_2(n) + 1$ least significant bits of its hash, while
the fingerprint consists of the remaining more significant bits. Therefore,
this process repurposes the least significant bit of each fingerprint to become
the most significant bit of the new canonical slot address. It also increments
the age counter of the old fingerprints, meaning that the resulting fluid
fingerprints retain a length of $f$ bits.
Fig.~\ref{fig:expandable_Memento_filter_example}-(B) shows an example of such
an expansion, followed by the insertion of a new key $x$ with a prefix
hash of $h(p(x))=$ \textbf{110}000 and memento $m(x)=$ \textit{0}. Analogously,
Memento filter contracts by halving the number of slots and transitioning a bit
from the addresses to the fingerprints.

\textbf{Insertions.} 
To maintain a stable FPR, new insertions are made with full-length
fingerprints. That is, a key $x$ is inserted by first finding a keepsake box
with a \emph{full-length matching fingerprint}. If one exists, $m(x)$ is added
to it. Otherwise, Memento filter creates a new keepsake box with a full-length
fingerprint for $x$ to minimize the FPR.
Fig.~\ref{fig:expandable_Memento_filter_example}-(B) shows an example. Even
though the new key $x$ has a partially matching fingerprint with the migrated
keepsake box in Slot 000, it manifests as a separate keepsake box. Note that an
$f$-bit fluid fingerprint can represent fingerprints of length at most~$f - 1$
bits. Thus, each slot and fluid fingerprint of the filter must be one bit wider
to maintain the same fingerprint length and FPR as a standard Memento filter.

\textbf{Deletions.} 
When deleting a key $y$, Memento filter removes a memento equal to $m(y)$ from
the keepsake box with the \emph{longest matching fingerprint}. The reason is
that deleting a memento with a shorter associated fingerprint may cause false
negatives, as it may have resulted from a hash collision with a different
keepsake box.

\textbf{Queries.}
Queries are handled as described in Section~\ref{sec:memento_filter}, but
Memento filter must probe all keepsake boxes with matching fluid fingerprints
for potential mementos. For example, in
\mbox{Fig.~\ref{fig:expandable_Memento_filter_example}-(B)}, a point query with
a prefix hash of \textbf{110}000 must check both keepsake boxes at Slots 000
and 001, which have fingerprints \textbf{0} and \textbf{110}. This does not
damage query performance, as partially matching fingerprints are rare, and
memory is still accessed sequentially.

\textbf{Unbounded Expansions.}
These methods allow Memento filter to expand up to $f - 1$ times, implying that
it can grow by a factor of up to $2^{f - 1}$. For typical fluid fingerprint
lengths such as $f = 11$ bits, this translates to Memento filter expanding up
to $2^{10} = 1024$ times its original size, which is sufficient for many
applications.

However, Memento filter fails to expand more than $f - 1$ times, as the oldest
fingerprints run out of bits to sacrifice. Memento filter can overcome this by
applying InfiniFilter~\cite{InfiniFilter}'s chaining method. Concretely, when a
keepsake box's fingerprint is depleted, it is removed from the filter and
inserted into a smaller, secondary Memento filter, where the hash is long
enough to create a full-length fingerprint. The secondary filter expands until
its fingerprints run out of bits, at which point it is added to a chain of
filters, and a new secondary filter is created. New insertions always go to the
main filter, but deletions and queries must probe all the filters. 

Speeding up queries and deletions in this case is an interesting direction for
future work. One approach may be to duplicate exhausted fingerprints across the
slots that could correspond to it, similarly to what Aleph Filter
proposes~\cite{AlephFilter}.

\textbf{Rejuvenation.}
Memento filter employs InfiniFilter's ``rejuvenation" operation. Here, a
fingerprint is lengthened during a positive query, whereby the application
regains access to the original key and can thus rehash it to derive a longer
fingerprint, improving the filter's FPR while also delaying the creation of
secondary filters.

\section{Theoretical Analysis}\label{sec:theoretical_analysis}
We show that Memento filter is close to space optimal and that its operation
costs are low. The last row of Table~\ref{tab:existing_method_stats} summarizes
the results of this section.

\textbf{False Positive Rate.}
We begin by answering the following question: \emph{Given a prefix $p^*$, what
is the probability $P^*$ that there is a different partition with some prefix
$p \neq p^*$ with a matching keepsake box?} The answer to this question is a
conservative upper bound on the FPR for both point and range queries, as it
does not take into account the \emph{non-robust} filtering the mementos
provide.

There are at most $\frac{N}{\ell}$ such partitions with prefix $p \neq p^*$,
where $N$ is the number of keys inserted into the filter and $\ell$ is the
average partition size. For $p$ to have a matching keepsake box with~$p^*$, it
must have the same canonical slot, i.e., $h(p) = h(p^*)$. If it shares the same
canonical slot, it must also share the same fingerprint, i.e., $h_f(p) =
h_f(p^*)$. The former has a probability of $\frac{1}{n}$, and the latter has a
probability of $2^{-f}$. Thus, since these events are independent, the
probability that $p$ and $p^*$ have matching keepsake boxes is $\frac{1}{n}
\cdot 2^{-f}$. Then, via a union bound on the total possible partitions with
prefix $p \neq p^*$, one can see that $P^* \leq \frac{N}{\ell} \cdot
\frac{1}{n} \cdot 2^{-f} = \frac{\alpha}{\ell} \cdot 2^{-f}$.

A negative point query for key $q$ can only result in a false positive when
there is some partition with prefix $p \neq p(q)$ with the same fingerprint and
a memento equal to $m(q)$. The probability $\epsilon_p$ of this event is upper
bounded by $P^* = \frac{\alpha}{\ell} \cdot 2^{-f}$, since $P^*$ only accounts
for the existence of such a partition. Depending on the key distribution, the
mementos may provide much better filtering and improve the FPR by a factor of
at best~$2^{-r}$.

Analogously, a negative range query $q=[q_l, q_r]$ can only result in a false
positive if there is some partition with a prefix $p \neq p(q_l), p(q_r)$ that
has a memento in the target memento ranges. Such a partition exists with
probability $\epsilon_r$ at most $2P^* = \frac{\alpha}{\ell} \cdot 2^{1 - f}$,
due to a union bound applied to $p(q_l)$ and $p(q_r)$. The mementos may further
improve the FPR by a constant factor. Thus, the overall FPR $\epsilon \leq
\text{max}(\epsilon_r, \epsilon_p)$ is at most $2P^* = \frac{\alpha}{\ell}
\cdot 2^{1 - f}$.

Using this result, one can follow a similar analysis to
InfiniFilter~\cite{InfiniFilter} and derive a bound of $\epsilon \leq (E + 2)
\cdot \frac{\alpha}{\ell} \cdot 2^{-f}$ for the FPR of an expandable Memento
filter, where $E \leq \log_2(N)$ is the number of expansions the filter has
undergone. 

\textbf{Expected Cluster Length.}
Let $\beta(l)$ be the length of the encoding of a keepsake box with size $l$,
measured in slots. We prove the following bound on the expected cluster length
$\mathbb{E}[|\mathcal{C}|]$ (see
Appendix~\ref{appendix:expected_cluster_size}):
\begin{theorem}
    \label{theorem:expected_cluster_size}
    $\mathbb{E}[|\mathcal{C}|] \leq \frac{\alpha \gamma(\ell)}{(1 - e^{-\alpha
    / \ell}) \cdot (1 - \alpha) \cdot (\gamma(\ell) - \alpha)}$, where
    $\gamma(\ell) = \frac{\ell}{\beta(\ell)}$.
\end{theorem}

Since $\alpha \leq 0.95$ and $1 \leq \gamma(\ell) \leq 1 + f/r$, the expected
cluster size will be $O(\ell)$. This further implies that there is an $O(1)$
number of keepsake boxes in the average cluster. In practice, we have found
$\ell$ to be close to one, implying that Memento filter will have a constant
expected cluster length. Theorem~\ref{theorem:expected_cluster_size} further
demostrates the excellent scalability of Memento filter with extreme dataset
skew. That is, when $\ell$ is small, $\gamma(\ell)$ will be close to one.
Therefore, assuming $\alpha = 0.95$, we have that $\mathbb{E}[|\mathcal{C}|]
\leq \frac{\alpha}{(1 - e^{-\alpha}) \cdot (1 - \alpha)^2} \approx 619.64$,
which matches a standard RSQF. However, as $\ell$ increases, $\gamma(\ell)$
tends to $1 + f/r$, which is typically at least $2$. In this case,
$\mathbb{E}[|\mathcal{C}|] \lessapprox \frac{2}{1 - \alpha} \cdot \ell$,
meaning that clusters remain as small as possible. As an example, assuming $f =
r$, $\alpha = 0.95$, and $\ell = 7$, we have that $\mathbb{E}[|\mathcal{C}|]
\lessapprox 413.77$, which improves upon an RSQF. 

\textbf{Performance.}
An insertion into Memento filter locates the target keepsake box and adds the
new key's memento to it. In the worst case, this operation will read and shift
the entire cluster of the keepsake box. Since the expected cluster length is
$O(\ell)$, an insertion also has an expected running time of $O(\ell)$. A
deletion follows an analogous procedure and thus has an $O(\ell)$ expected
execution time.

A point query locates the appropriate keepsake box by skipping a constant
number of keepsake boxes in the cluster and searches for the target memento,
requiring a total of $O(\log_2 \ell)$ operations. Range queries access either
one or two keepsake boxes. The former case's analysis is identical to the case
of point queries. In the latter case, both keepsake box lookups take $O(1)$
operations, and only the largest and smallest mementos are ever accessed for
each, which requires $O(1)$ time since they are stored near the keepsake box's
fingerprint. Thus, the cost of a range query is $O(\log_2 \ell)$.

Notice that each keepsake box lookup entails only a single random cache miss on
average. Probing a keepsake box is done using the already cached memory
segments and incurs no further cache misses. Since clusters and keepsake boxes
are arranged sequentially, when a cluster becomes too large to fit in a cache
line, the resulting extraneous memory accesses and cache misses are all
sequential, thus taking full advantage of the hardware prefetcher.

In conclusion, on average, Memento filter will incur a single cache miss for
insertions, deletions, and point queries, while range queries are serviced with
up to two random cache misses. 

\textbf{Memory Footprint.}
Each slot in Memento filter is $f + r$ bits long. With the FPR analysis in
mind, Memento filter can guarantee an FPR of $\epsilon$ with a fingerprint
length of $f = 1 + \log_2 \frac{1}{\epsilon}$. Furthermore, to support range
queries of length $R$, $r$ must be at least $\log_2 R$ bits. Taking into
account the metadata overhead of the RSQF, a Memento filter with a load factor
of $\alpha$ will have a memory footprint of $\frac{1}{\alpha} (3.125 + \log_2
\frac{R}{\epsilon})$. In the case of an expandable Memento filter, since each
slot is one bit wider to accommodate the unary age counter, the memory
footprint becomes $\frac{1}{\alpha} (4.125 + \log_2 \frac{R}{\epsilon})$.

\section{Evaluation}
We compare Memento filter to existing range filters in a standalone setting in
Section~\ref{sec:standalone_evaluation} In Section~\ref{sec:b_tree_evaluation},
we provide experimental results from our integration of Memento filter with
WiredTiger, a B-tree based key-value store. We utilize Grafite's benchmark
template for our evaluations~\cite{Grafite}.

\textbf{Platform.}
We use a Fedora 39 machine with a single Intel Xeon w7-2495X processor (4.8
GHz) with 24 cores and 48 hyperthreads. It has 64 GBs of main memory, a 45 MB
L3 cache, a 48 MB L2 cache, and a 1920 kB L1 cache. It also has two SK Hynix
512 GB PC611 M.2 2280 80mm SSDs, with a sequential read/write performance of up
to 3400/2700 MBps and random read/writes of up to 440K/440K IOps. These SSDs
are used in the B-Tree experiments only.

\subsection{Standalone Evaluation}\label{sec:standalone_evaluation}

\begin{figure*}
    \centering
    \includegraphics[width=\textwidth]{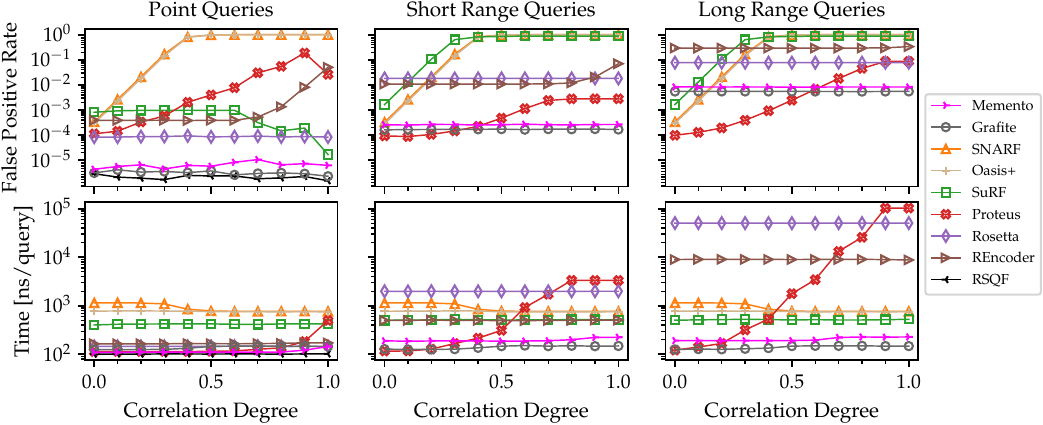}
    \caption{Most filters (SNARF, Oasis+, SuRF, Proteus, REncoder) exhibit
    worse FPRs as the workload becomes increasingly correlated and are thus not
    robust. Some of these filters (SNARF, Proteus) also have varying query
    times with different correlation degrees. Only Rosetta, Grafite, and
    Memento filter have a robust FPR guarantee and at the same time, stable
    query costs.}
    \label{fig:correlation_test}
\end{figure*}

\textbf{Baselines.}
We conduct experiments over both static and dynamic data. In the static
setting, we compare Memento filter with SuRF~\cite{SuRF},
Rosetta~\cite{Rosetta}, REncoder~\cite{REncoder, REncoder_Journal},
Proteus~\cite{Proteus}, SNARF~\cite{SNARF}, Oasis+~\cite{Oasis}, and
Grafite~\cite{Grafite}. We do not include bloomRF~\cite{bloomRF} as a baseline
as it is closed-source. In the dynamic setting, we only compare the expandable
version of Memento filter with Rosetta, REncoder, and SNARF, as other filters
do not support incremental updates. We implement Memento filter in \verb|C| and
use the open-source \verb|C/C++| implementations of the baselines. All filters
are compiled with \verb|gcc-13|.

We employ the original key suffixes in the leaves of SuRF when considering
range query workloads to allow for comparing query end-points at the leaves,
and use hash suffixes when considering point query workloads. We allow Rosetta
and Proteus to auto-tune their memory allocation with a query sample,
showcasing their best performance. We tune Memento filter with a memento size
$r$ based on the maximum query size in the workload.

\begin{figure*}
    \centering
    \begin{tabular}{cc}
        \includegraphics[width=0.635\textwidth]{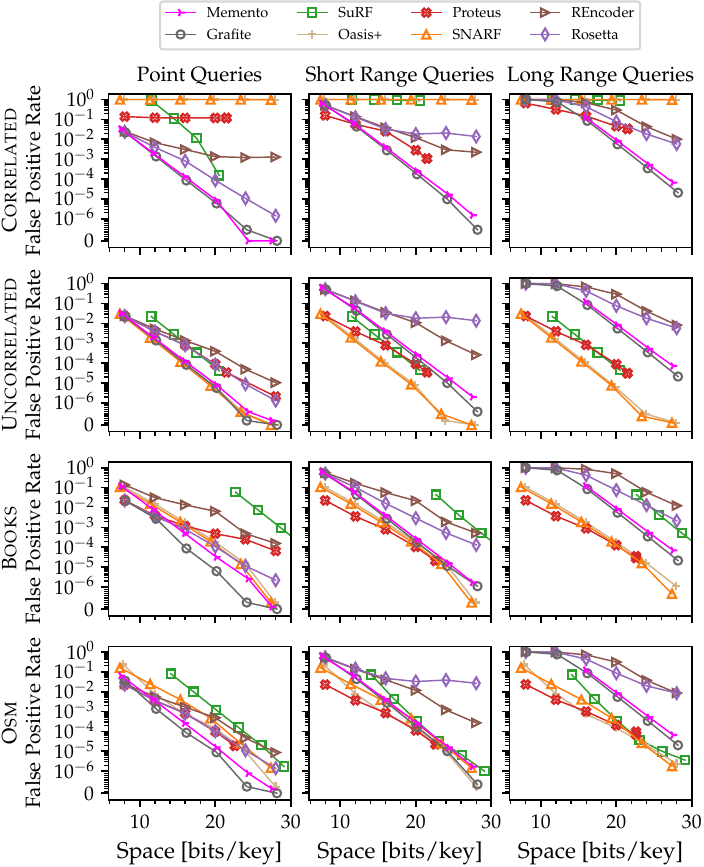} & 
        \setlength\extrarowheight{-1.85pt}
        \scriptsize
        \begin{tabular}[b]{cc}
            \toprule
            \textbf{Competitor} & \makecell[c]{\textbf{Avg. Query Time [ns]} \\ \textbf{(vs. Memento)}} \\
            \midrule
            Memento & 176 (1.0$\times$) \\
            Grafite & 147 (0.84$\times$) \\
            SuRF & 477 (2.71$\times$) \\
            Oasis+ & 748 (4.25$\times$) \\
            SNARF & 760 (4.32$\times$) \\
            REncoder & 1461 (8.3$\times$) \\
            Rosetta & 12106 (68.78$\times$) \\
            Proteus & 14896 (84.64$\times$) \\
            \midrule
            Memento & 160 (1.0$\times$) \\
            Proteus & 115 (0.72$\times$) \\
            Grafite & 130 (0.81$\times$) \\
            SuRF & 241 (1.51$\times$) \\
            Oasis+ & 649 (4.06$\times$) \\
            SNARF & 1028 (6.42$\times$) \\
            REncoder & 1459 (9.12$\times$) \\
            Rosetta & 12068 (75.42$\times$) \\
            \midrule
            Memento & 240 (1.0$\times$) \\
            Grafite & 128 (0.53$\times$) \\
            Proteus & 141 (0.59$\times$) \\
            SuRF & 292 (1.22$\times$) \\
            Oasis+ & 731 (3.05$\times$) \\
            SNARF & 1078 (4.49$\times$) \\
            REncoder & 1208 (5.03$\times$) \\
            Rosetta & 11855 (49.4$\times$) \\
            \midrule
            Memento & 157 (1.0$\times$) \\
            Grafite & 129 (0.82$\times$) \\
            Proteus & 153 (0.97$\times$) \\
            SuRF & 345 (2.2$\times$) \\
            Oasis+ & 817 (5.2$\times$) \\
            REncoder & 1440 (9.17$\times$) \\
            SNARF & 1620 (10.32$\times$) \\
            Rosetta & 12036 (76.66$\times$) \\
            \bottomrule
            \vspace{14pt}
        \end{tabular}
    \end{tabular}
    \caption{Memento filter and Grafite provide the best filtering in the case
    of correlated workloads, the best point filtering in general, and the
    fastest overall query speed. Even though Memento filter and Grafite are
    competitive with the state of the art on real workloads, they provide less
    filtering compared to their heuristic counterparts when considering an
    uncorrelated workload due to the strong filtering guarantees they
    provide.}
    \label{fig:fpr_test}
\end{figure*}

\textbf{Datasets.}
We conduct our experiments with the same synthetic and real-world
datasets~\cite{SOSD-vldb, SOSD-neurips} used in previous range filter
evaluations~\cite{SuRF, Rosetta, REncoder, REncoder_Journal, bloomRF, Proteus,
SNARF, Grafite}:
\begin{itemize}
    \item \textsc{Uniform}: 200M 64-bit integers chosen uniformly at random.
    \item \textsc{Normal}: 200M 64-bit integers sampled from $\mathcal{N}(2^{63},
        0.1 \cdot 2^{63})$.
    \item \textsc{Books}: Amazon booksale popularity for 200M books. 
    \item \textsc{OSM}: 200M location coordinates from the Open Street Map.
\end{itemize}

\textbf{Static Workloads.}
Following existing works~\cite{SuRF, Rosetta, REncoder, REncoder_Journal,
bloomRF, Proteus, SNARF, Grafite}, we create a set of 10M range queries of the
form $[x, x + R - 1]$, where $x$ is a key from the key universe and $R$ is the
range query length. We run separate workloads with point queries ($R=1$), short
range queries ($R=2^5=32$) and long range queries ($R=2^{10}=1024$). We choose
the starting point $x$ of the queries in one of three ways:
\begin{itemize}
    \item \textsc{Uncorrelated}: $x$ is chosen uniformly at random.
    \item \textsc{Correlated}: $x$ is chosen by first considering a randomly
        chosen key $k$ from the dataset, and sampling from the range $[k, k +
        2^{30 \cdot (1-D)}]$, where $D$ is the correlation degree of the
        workload. By default, we set $D=0.8$.
    \item \textsc{Real}: $x$ is sampled and removed from the underlying
        dataset.
\end{itemize}
In all these workloads, we only consider empty query ranges, allowing us to
measure the FPR as the ratio of positive results to the query batch size. We
also provide a separate experiment detailing filter throughput for positive
queries. We only consider the filter query times in our standalone experiments
and not the time required to access a slower storage medium.

\textbf{Experiment 1: Robustness to Correlated Workloads.}
We evaluate the robustness of the range filters by using the static
\textsc{Uniform} dataset and a \textsc{Correlated} query workload with a
varying correlation degree from 0 to 1. All filters are assigned a memory
budget of 20 bits per key. The first row of Fig.~\ref{fig:correlation_test}
shows that only Rosetta, Grafite, and Memento filter are unaffected by workload
correlation and are thus robust. Both Memento filter and Grafite have better
FPRs than Rosetta by up to two orders of magnitude. As shown, Memento filter
approximately matches the FPR of Grafite. All other filters exhibit increasing
FPRs with more correlation.

Notice that, when considering point queries, SuRF's FPR actually decreases with
higher correlation degrees. This is due to SuRF comparing key hashes in this
case, which provides much better filtering when the workload is heavily
correlated.

The second row of Fig.~\ref{fig:correlation_test} shows that Grafite and
Memento filter are the most efficient range filters in terms of query speed,
improving upon all other filters by a factor of at least $4\times$. Memento
filter provides faster point queries than Grafite by 20\%, while closely
matching Grafite's performance in servicing range queries. 

We have included evaluation results for a vanilla RSQF with the same memory
footprint in the point query column of Fig.~\ref{fig:correlation_test}. As
shown, Memento filter achieves an FPR competitive with a standard RSQF while
adding negligible overhead to queries. 

\textbf{Experiment 2: FPR vs. Memory Tradeoff.}
Fig.~\ref{fig:fpr_test} shows an FPR comparison of all range filters on
synthetic and real-world data. In the synthetic case, we consider the
\textsc{Uniform} dataset and execute both \textsc{Correlated} and
\textsc{Uncorrelated} workloads. For the real workloads, we use the
\textsc{Books} and \textsc{OSM} datasets, along with \textsc{Real} query
workloads. Each row of Fig.~\ref{fig:fpr_test} provides experiment results for
a single dataset and workload with varying range sizes, as well as query speed
statistics averaged over all range query sizes. 

We only provide partial graphs for Memento filter when considering long range
queries, since Memento filter requires at least 12 bits to store metadata and a
large enough memento in this case. Notice that all other robust range filters
exhibit an FPR of 1 below this space threshold, only wasting memory.

As established before, Grafite and Memento filter have the best FPR when the
workload is correlated. Memento filter is competitive with Grafite with only a
$1.5 \times$ gap in FPR and provides up to 5 orders of magnitude better FPR
than Rosetta. Furthermore, Memento filter and Grafite provide the best point
filtering across all datasets. However, as the range sizes increase in
non-correlated workloads, robust range filters provide less filtering than
their heuristic competitors due to their strong FPR guarantees.

In terms of query speed, Grafite and Memento filter provide the best overall
performance. Memento filter is slightly slower than Grafite but provides
dynamic insertions and deletions in exchange. Even though Proteus is faster
than Memento filter and Grafite in the last three rows of
Fig.~\ref{fig:fpr_test}, it does not guarantee a robust FPR and is slower when
considering correlated workloads.

\begin{figure*}
    \centering
    \includegraphics[width=0.9\textwidth]{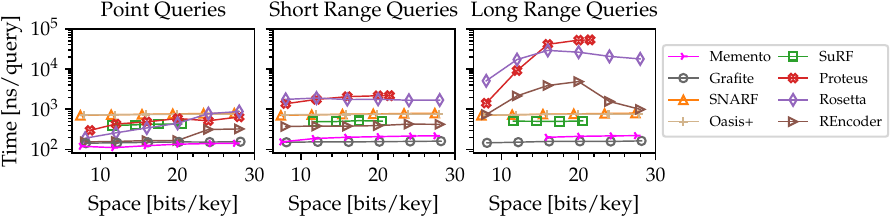}
    \caption{Grafite and Memento filter provide the best and most stable
    performance when non-empty queries are involved.}
    \label{fig:non_empty_test}
    \vspace{-0.5cm}
\end{figure*}

\textbf{Experiment 3: Non-Empty Query Performance.}
Although filters are typically used to reduce slower media accesses, such as
network calls and disk reads, they must minimize their added CPU overhead for
non-empty queries as well. We thus benchmark filters throughputs on non-empty
queries in Fig.~\ref{fig:non_empty_test} by using the \textsc{Uniform} dataset
and creating query ranges of the form $[x, x + R - 1]$, where $x$ is sampled
from $[k - L + 1, k]$ for a randomly chosen key $k$ in the dataset. We also
experimented with the \textsc{Normal} and \textsc{Real} datasets, but omit
their results as the best filters and their performance remains the same. The
results show that Grafite and Memento filter are the fastest to process
positive range queries and provide stable performance with varying memory
budgets. Memento filter matches Grafite's performance in processing range
queries and has faster point queries by up to 38\%.

\begin{figure}
    \centering
    \begin{minipage}{.425\textwidth}
        \centering
        \includegraphics[width=\columnwidth]{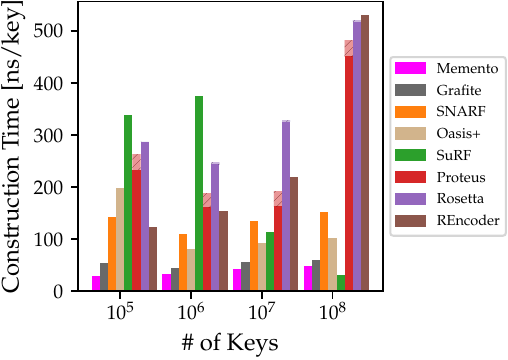}
        \caption{Memento filter provides the best construction time in almost all
        scenarios.}
        \label{fig:construction_time_test}
    \end{minipage}%
    \hspace{4pt}
    \begin{minipage}{.55\textwidth}
        \centering
        \includegraphics[width=\columnwidth]{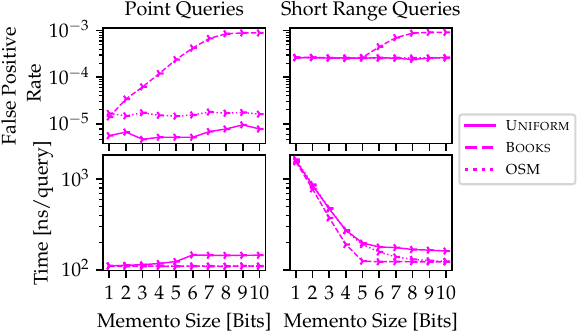}
        \caption{Memento filter maintains its FPR guarantee if the memento size
        is too small but exhibits slower queries. If the memento size is too
        large, Memento filter maintains its original query speed, but its FPR
        may suffer.}
        \label{fig:vary_memento_test}
    \end{minipage}
\end{figure}

\textbf{Experiment 4: Construction Time.}
Fig.~\ref{fig:construction_time_test} compares the construction times of all
range filters with varying dataset sizes. Since the choice of dataset does not
influence the construction times of the filters, we use the \textsc{Uniform}
dataset. We report construction time averages over various memory budgets. The
light colors of Fig.~\ref{fig:construction_time_test} used for Rosetta and
Proteus indicate the impact of their tuning processes, evaluated with an
\textsc{Uncorrelated} query workload with $\frac{N}{10}$ queries, where $N$ is
the number of keys in the dataset.

Memento filter achieves the best construction time in almost all cases, beating
its closest competitor by 20\%. Memento filter's bulk loading algorithm can be
further optimized by using a multi-threaded sorting algorithm to sort the key
hashes. Moreover, since Memento filter is a dynamic filter, it can also be
constructed by streaming the keys. Therefore, Memento filter can be constructed
in a single pass of the data without the need for sorting, providing
significant speedup when the dataset is too large to fit in memory.

\textbf{Experiment 5: Memento Size Choice.}
Accurately estimating the maximum range query length $R$ is integral to Memento
filter's performance, as the memento size $r$ is chosen to be $\lceil \log_2 R
\rceil$. In practice, however, users may err in estimating $R$ and thus in
setting $r$. Fig.~\ref{fig:vary_memento_test} shows how Memento filter's FPR
and query speed vary for different memento size configurations under a memory
budget of 20 bits per key. Here, longer mementos imply shorter fingerprints and
vice-versa. We consider various datasets and issue queries from a correlated
workload ($D = 0.8$). We also experimented with uncorrelated workloads but have
omitted the results, as they are at least as good as the correlated case. The
first column considers point queries, while the second column showcases short
range queries with $R = 32$. Thus, the optimal memento length $r^*$ in the
first column is $r^* = 1$, while for the second column it is $r^* = 5$.

Fig.~\ref{fig:vary_memento_test} shows that Memento filter's FPR does not
deviate from the optimal as long as $r \leq r^*$. However, its query time
worsens by a factor of $2 ^ {r^* - r}$ due to the extra lookups, which is
proportional to the user's estimation error of $R$. In contrast, if $r > r^*$,
Memento filter may exhibit a higher FPR depending on the dataset. The reason is
that the robust filtering provided by the fingerprints is replaced with the
non-robust filtering provided by the mementos. Fig.~\ref{fig:vary_memento_test}
shows that the FPR does not worsen indefinitely and saturates at a
dataset-dependent value. Furthermore, Memento filter maintains its optimal
query speed, except for point queries, where it incurs a slight slowdown due to
the extra memento comparisons.

We advise practitioners to estimate a lower bound of $R$ as close as possible
to the actual value, preserving the excellent FPR guarantee of Memento filter
in exchange for slightly slower queries. As filters typically use 1-3 bytes per
key in practice, one cannot construct a robust range filter for large $R$ due
to the information-theoretic lower bound. For moderate $R$, having one-byte
fingerprints and one to two-byte mementos is common.

\begin{figure*}
    \centering
    \includegraphics[width=\textwidth]{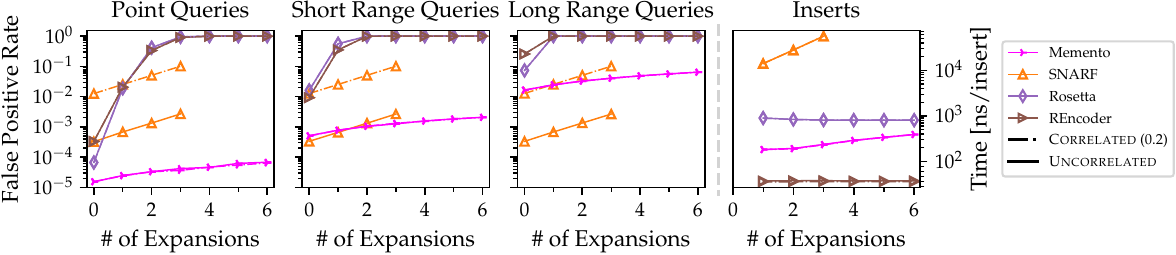}
    \caption{Memento filter is the only filter that maintains fast insertion
    times and a low false positive rate across expansions.}
    \label{fig:expandability_test}
\end{figure*}

\textbf{Dynamic Workloads.}
We consider the \textsc{Uniform} dataset and construct the filter on a random
$\frac{1}{64}$ fraction of the data. We then insert keys into the filter one by
one until an \emph{Expansion} occurs, i.e., the digested data size doubles. We
continue this process until the entire dataset is inserted into the filter.
After each expansion, we measure performance statistics by running 10M queries
from both \textsc{Uncorrelated} and \textsc{Correlated} workloads.

\textbf{Experiment 6: Expanding Datasets.}
We compare Memento filter with Rosetta, REncoder, and SNARF (the only other
filters supporting incremental insertions) in a dynamic setting.
Fig.~\ref{fig:expandability_test} plots the FPR and insertion times of these
filters across expansions. All filters are constrained to a memory budget of 20
bits per key. 

Even though the filters have a similar initial FPR, only Memento filter
maintains its FPR guarantee across expansions. It also maintains its excellent
insertion speed. Rosetta and REncoder provide no filtering after just three
expansions, while SNARF fails to accommodate new insertions efficiently. We do
not plot all of SNARF's performance metrics, as it takes over 5 hours to expand
after the third expansion. Even though SNARF still provides better filtering
than Memento filter in the face of long and mixed \textsc{Uncorrelated} range
queries, it is worse in all cases as soon as the workload becomes slightly
correlated (even with a correlation degree of $0.2$).

Memento filter provides better insertion times than Rosetta but is slower than
REncoder. Moreover, its insertion throughput is decreasing slightly. This is
due to Memento filter expanding when the dataset size doubles, causing a
smaller fraction of the filter to fit in the higher levels of cache in exchange
for maintaining its FPR.

It is worth noting that only Memento filter is compatible with InfiniFilter's
techniques, as it is a tabular filter. All other range filters utilize bitmaps
and Bloom filters in their structures, which makes them unable to expand
without rescanning the data from storage. \\[-0.7cm]

\subsection{B-Tree Evaluation}\label{sec:b_tree_evaluation}
B-Trees~\cite{BTree, B+Tree} are the de facto standard for file organization
and indexing tasks. These structures are search trees that minimize data
movement -- the main bottleneck of database systems. Similarly to a
binary search tree, the internal nodes of a B-Tree partition the search space
of the key set into $B$ partitions, where $B$ is dictated by the data movement
granule the system offers and the key size. The leaves of the tree contain
the entries themselves in sorted order. As entries are added and removed from
the tree, it rebalances to maintain robust performance.

Databases use this data structure to achieve efficient random access to keys.
They can also scan specific ranges of the entries, as the tree is
order-preserving. B-Trees are further optimized using \emph{Buffer Pools},
which cache frequently accessed nodes in main memory to reduce data
movement and access latency~\cite{BufferPools}.

B-Trees are ubiquitous in many industrial applications. For example,
MongoDB~\cite{MongoDB}, a popular document database, uses a B-Tree-based
key-value store called WiredTiger~\cite{WiredTiger} as its backend. However,
B-Trees are often subject to workloads with many empty short range queries,
comprising up to 50\% of their queries. This is observed in several database
applications, such as social graph analytics~\cite{TAO, LinkBench}. Thus,
B-Trees are a prime example of an application that can significantly benefit
from a dynamic range filter.

Due to WiredTiger's widespread industrial use, we integrate the expandable
version of Memento filter with it. We create a single instance of our filter,
which is constructed over the entire data. To the best of our knowledge, we are
the first to integrate a range filter with a B-Tree, a feat previously
impossible due to the dynamicity of B-Trees which necessitates a
dynamic/expandable range filter.

\textbf{Datasets.}
We conduct our evaluation with subsets of size 100M  from the \textsc{Uniform},
\textsc{Normal}, and \textsc{Books} datasets. In all cases, we store randomly
generated 504-byte values in the B-Tree to make for 512-byte key-value pairs.

\textbf{Workloads.}
We employ a workload similar to the dynamic workload described in
Section~\ref{sec:standalone_evaluation}, but initialize the system on a random
$\frac{1}{8}$ fraction of the dataset considered. To measure performance
statistics, we run 10M mixed range queries of length $1 \leq R \leq 32$, where
the left end-point is sampled from the same distribution of the keys, i.e.,
\textsc{Uniform}, \textsc{Normal}, and \textsc{Real}. We vary the percentage of
non-empty queries in the workload to provide a clear overall picture of the
system's performance in different scenarios.

\textbf{Baselines.}
Since Memento filter is the only dynamic and expandable range filter, we only
compare it with a standard instance of WiredTiger. This instance will use all
of its allocated main memory for a buffer pool, allowing it to cache many of
the B-Tree's nodes. When integrating the expandable version of Memento filter,
we reallocate some of the buffer pool's memory for a Memento filter with a
memory budget of 15 bits per key to draw a fair comparison. We allocate a total
memory budget equivalent to 2\% of the current dataset size to both instances. 

\begin{figure}
    \centering
    \includegraphics[width=0.68\columnwidth]{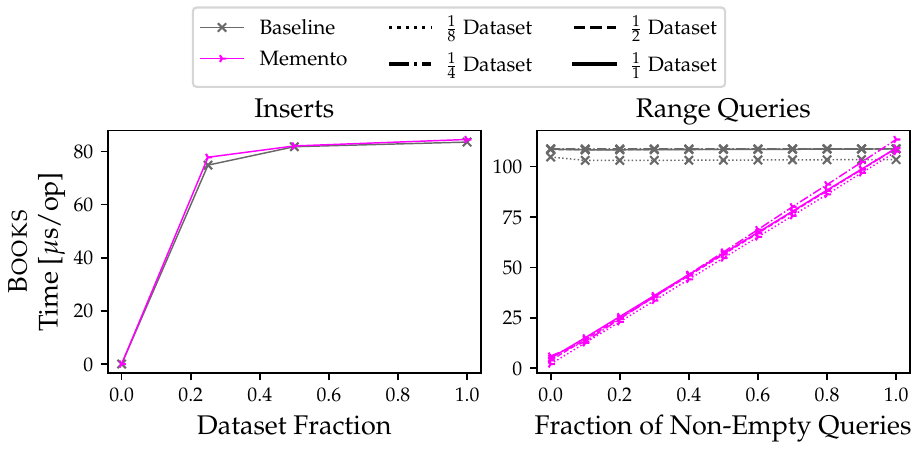}
    \caption{Memento filter significantly improves the query throughput of
    WiredTiger when empty queries are present while maintaining the same
    insertion performance.}
    \label{fig:b_tree_test}
\end{figure}

\textbf{Experiment 7: B-Tree Performance.}
The results of Fig.~\ref{fig:b_tree_test} show that WiredTiger benefits immensely from
Memento filter at all data sizes, achieving faster query processing by $1.9
\times$ when 50\% of the queries are empty. Note that we only show one plot, as
all datasets have similar results. WiredTiger also maintains its query
throughput when all queries are non-empty. Furthermore, Memento filter does not
significantly affect the insertion times of WiredTiger in both workloads,
incurring a minor overhead of $\approx 2.5\%$. Thus, trading off buffer pool
memory for a Memento filter in workloads with many empty queries may
significantly improve the system's overall performance.

\section{Conclusion}
We introduced Memento filter, the first dynamic range filter with fast
operations and a robust false positive rate guarantee. By encoding keepsake
boxes in an RSQF, Memento filter achieved FPRs and performance on par with the
state of the art. It further achieved expandability by employing
variable-length fingerprints. We argued that Memento filter is the only
practical dynamic range filter, and solidified our claim by integrating it with
WiredTiger, showing that it significantly boosts range query performance while
not hindering insertions. Further exploring the tradeoffs of using Memento
filter in a fully functional system, the tradeoffs of partitioning it into
smaller filters, and its cacheability are intriguing directions for future
work. Additionally, exploring other design choices, such as adaptively
partitioning the key universe, will be fruitful.

\begin{acks}
    We thank the reviewers for their insightful comments. This research was
    supported by the NSERC grant \#RGPIN-2023-03580.
\end{acks}

\bibliographystyle{ACM-Reference-Format}
\bibliography{memento}

\pagebreak
\appendix
\section{Expected Cluster Size}\label{appendix:expected_cluster_size}
\begin{proof}
Consider a randomly chosen cluster $\mathcal{C}$ starting in Slot $p$. We
inductively define two series of random variables $X_i$ and $Y_i$ based on
$\mathcal{C}$ as follows:
\begin{itemize}
    \item $X_1$ represents the number of partitions in the key universe
        mapping to Slot $p$.

    \item $Y_1$ represents the number of keys in the partitions considered in
        $X_1$.

    \item $X_i$ represents the number of partitions in the key universe mapping
        to any of the slots in the range $[p+\sum_{j=1}^{i-1} Z_j,
        p-1+\sum_{j=1}^i Z_j]$, where $Z_j$ is the number of slots filled by
        the keys considered in $Y_j$. Since $Z_j \leq Y_j$ in Memento filter
        due to the keepsake box encoding scheme, one can simplify the
        definition of $X_i$ by instead considering partitions mapping to the
        range $[p+\sum_{j=1}^{i-1} Y_j, p-1+\sum_{j=1}^i Y_j]$ containing more
        slots, slightly overestimating $X_i$.

    \item $Y_i$ represents the number of keys in the partitions considered in
        $X_i$.
\end{itemize}
Observe that $\mathbb{E}[|\mathcal{C}|] \leq \mathbb{E}[\sum_{i=1}^{N/\ell}
Z_i] \leq \mathbb{E}[\sum_{i=1}^{N/\ell} Y_i] = \sum_{i=1}^{N/\ell}
\mathbb{E}[Y_i]$. We thus bound the $\mathbb{E}[Y_i]$s to bound
$\mathbb{E}[|\mathcal{C}|]$. First notice that
\vspace*{-0.17cm}
\begin{equation}
    \mathbb{E}[Y_i] = \mathbb{E}[\mathbb{E}[Y_i | X_1,\dots,X_i,Y_1,\dots,Y_{i-1}]]
    = \mathbb{E}\left[\frac{N - \sum_{j=1}^{i-1} Y_j}{N/\ell - \sum_{j=1}^{i-1} X_j} \cdot X_{i-1} \right]
    \label{equation:Y}
\end{equation}
since all partitions in $X_i$ map to the desired range equi-probably and their
expected total number of keys $\mathbb{E}[Y_i]$ equals their average size times 
the number of partitions $X_i$. Applying the Poisson approximation to the balls
and bins problem with partitions as balls and slots as bins, we get
\begin{equation}
    \mathbb{E}[X_i | X_1,\dots,X_{i-1},Y_1,\dots,Y_{i-1}]
    = \frac{N/\ell - \sum_{j=1}^{i-1} X_j}{n - \sum_{j=1}^{i-1} Z_j} \cdot Y_{i-1} \\
    \leq \frac{N/\ell - \sum_{j=1}^{i-1} Y_j}{n - \sum_{j=1}^{i-1} Y_j} \cdot Y_{i-1}.
    \label{equation:X}
\end{equation}
Putting equations~\ref{equation:Y} and \ref{equation:X} together, we conclude that 
\begin{align*}
    \mathbb{E}[Y_i] & \leq \mathbb{E}\left[\mathbb{E}\left[\frac{N - \sum_{j=1}^{i-1} Y_j}{N/\ell - \sum_{j=1}^{i-1} X_j} \cdot X_i | X_1,\dots,X_{i-1},Y_1,\dots,Y_{i-2}\right]\right] \\
    & \leq \mathbb{E}\left[\frac{N - \sum_{j=1}^{i-1} Y_j}{N/\ell - \sum_{j=1}^{i-1} X_j} \cdot \frac{N/\ell - \sum_{j=1}^{i-1} X_j}{n - \sum_{j=1}^{i-1} Y_j} \cdot Y_{i-1} \right] \\
    & \leq \mathbb{E}\left[\frac{N - \sum_{j=1}^{i-1} Y_j}{n - \sum_{j=1}^{i-1} Y_j} \cdot Y_{i-1} \right] \leq \mathbb{E}\left[\frac{N}{n} \cdot Y_{i-1} \right] = \alpha \cdot \mathbb{E}[Y_{i-1}],
\end{align*}
further implying $\mathbb{E}[Y_i] \leq \alpha^{i-1} \cdot \mathbb{E}[Y_1]$.
Moreover, since $\mathbb{E}[Y_1] = \mathbb{E}[\frac{N}{N/\ell} \cdot X_1] =
\ell \cdot  \mathbb{E}[X_1]$, we have that $\mathbb{E}[Y_i] \leq \alpha^{i-1}
\cdot \ell \cdot \mathbb{E}[X_1]$. Letting the random variable $W$ denote the
number of partitions mapped to Slot $p$ and denoting by $F$ the event where
Slot $p$ is the first slot in a cluster, we bound $\mathbb{E}[X_1]$ as 
\begin{align*}
    \mathbb{E}[X_1] & = \sum_{i=1}^{N/\ell} i \cdot \Pr(W=i | F) = \sum_{i=1}^{N/\ell} i \cdot \frac{\Pr(F | W=i) \cdot \Pr(W=i)}{\Pr(F)} 
    \leq \sum_{i=1}^{N/\ell} i \cdot \frac{\Pr(W=i)}{\Pr(F)} \\ 
    & \leq \frac{\alpha}{\ell \cdot \Pr(F)} \leq \frac{\alpha}{\ell \cdot \Pr(\text{Slot } p-1 \text{ empty} \wedge W>0)} \leq \frac{\alpha}{\ell \cdot (1 - \frac{\sum_i Z_i}{n}) \cdot (1-e^{-\alpha/\ell})} \\
    & \leq \frac{\alpha}{\ell \cdot (1 - \frac{N \cdot \beta(\ell)}{n \cdot \ell}) \cdot (1 - e^{-\alpha/\ell})} = \frac{\alpha \cdot \gamma(\ell)}{\ell \cdot (\gamma(\ell) - \alpha) \cdot (1 - e^{-\alpha/\ell})}.
\end{align*}
The last inequality uses Jensen's inequality in conjunction with the concavity
of $\beta(\cdot)$. Putting everything together results in
$\mathbb{E}[|\mathcal{C}|] \leq \sum_{i=1}^N \mathbb{E}[Y_i] \leq \mathbb{E}[X_1] \cdot \ell \cdot \sum_{i=1}^\infty \alpha^{i-1} = \frac{\alpha \cdot \gamma(\ell)}{(1-e^{-\alpha/\ell}) \cdot (\gamma(\ell) - \alpha) \cdot (1-\alpha)}$.
\end{proof}

\end{document}